\documentclass{aa}  

\usepackage{newtxtext}
\usepackage{color}
\usepackage{amsmath}
\usepackage{hyperref}
\usepackage{soul}
\usepackage[normalem]{ulem}
\usepackage{txfonts}

\begin{document}

   \title{Consequences of planetary migration on the minor bodies\\of the early solar system}

   \subtitle{}

   \author{S. Pirani\inst{1}
          \and
          A. Johansen\inst{1}
          \and
          B. Bitsch\inst{2}
          \and
          A. J. Mustill\inst{1}  
          \and
          D. Turrini\inst{3}
          }

   \institute{Lund Observatory, Department of Astronomy and Theoretical Physics, Lund University, Box 43, 22100 Lund, Sweden.\\
              \email{simona@astro.lu.se}
         \and
             Max-Planck-Institut f\"{u}r Astronomie, K\"{o}nigstuhl 17, 69117 Heidelberg, Germany.
         \and
             Institute for Space Astrophysics and Planetology INAF-IAPS, Via Fosso del Cavaliere 100, 00133 Rome, Italy.
             }

   \date{}

 
  \abstract
  {Pebble accretion is an efficient mechanism that is able to build up the core of the giant planets within the lifetime of the protoplanetary disc gas-phase. The core grows via this process until the protoplanet reaches its pebble isolation mass and starts to accrete gas. During the growth, the protoplanet undergoes a rapid, large-scale, inward migration due to the interactions with the gaseous protoplanetary disc.
  In this work, we have investigated how this early migration would have affected the minor body populations in our solar system. In particular, we focus on the Jupiter Trojan asteroids (bodies in the coorbital resonance 1:1 with Jupiter, librating around the L$_4$ and L$_5$ Lagrangian points called, respectively, the leading and the trailing swarm) and the Hilda asteroids. We characterised their orbital parameter distributions after the disc dispersal and their formation location and compare them to the same populations produced in a classical in situ growth model. 
We find that a massive and eccentric Hilda group is captured during the migration from a region between 5 and 8 au and subsequently depleted during the late instability of the giant planets. Our simulations also show that inward migration of the giant planets always produces a Jupiter Trojans' leading swarm more populated than the trailing one, with a ratio comparable to the current observed Trojan asymmetry ratio. The in situ formation of Jupiter, on the other hand, produces symmetric swarms. The reason for the asymmetry is the relative drift between the migrating planet and the particles in the coorbital resonance. The capture happens during the growth of Jupiter's core and Trojan asteroids are afterwards carried along during the giant planet's migration to their final orbits. The asymmetry and eccentricity of the captured Trojans correspond well to observations, but their inclinations are near zero and their total mass is three to four orders of magnitude higher than the current population. Future modelling will be needed to understand whether the dynamical evolution of the Trojans over billions of years will raise the inclinations and deplete the masses to observed values.}


   \maketitle
%

\section{Introduction}\label{sec:intro}

In the framework of the formation and growth of massive planets, the core accretion theory \citep{pollack96} assumes that the envelopes of gas giants are accreted after the formation of a massive core. However, to build up a core of roughly 10 M$_\oplus$ from small dust grains is not trivial. Millimetre-centimetre sized pebbles can grow from micron sized dust particles by coagulation or condensation \citep{brauer08,zsom10,birnstiel12,ros13,schoonenberg17}. 
When solid particles grow to millimetre-centimetre sizes, further growth by sticking collisions is inhibited by the `bouncing barrier' because they do not stick and accrete anymore, but start to bounce off each other \citep{guttler10,zsom10}.
Moreover, because solid particles are not supported by the pressure gradient as is the gas, their motion is affected by a headwind that causes them to lose angular momentum and hence to drift rapidly toward the central star \citep{weidenschilling77}. Differently sized particles drift with different velocities and this leads to high relative speeds and disruptive collisions and the establishment of a `fragmentation barrier' \citep{blum93}. Even without any `bouncing' or `fragmentation barrier', particles would quickly drift towards the central star before they can grow above a size where the radial drift is less efficient. Hence, growing grains must find a way to grow very quickly in order to overcome the `radial-drift barrier' \citep{weidenschilling77}, also called `metre-size barrier' because, in standard disc models, the drifting velocity has a maximum for metre-sized objects. 

Beyond the snow line, however, icy particles tend to form high porosity aggregates that are more resistant to collisions and are less affected by the radial drift \citep{wada09,okuzumi12}. \citet{windmark12a,windmark12b} show that if a small number of particles are lucky enough to overcome the `bouncing barrier' and grow up to centimetre sizes, they can sweep up the smaller particles still affected by the barrier and grow to planetesimals. 
Another possibility is the streaming instability: when solid particles cluster in the gas, the back reaction from the dust on the gas reduces the headwind felt by the pebbles locally, allowing the cluster to orbit faster and undergo less inward drift; the cluster can then be joined by other pebbles that are still drifting faster and the local solid density keeps increasing, further reducing radial drift, and leading to an exponential growth of the initial cluster; these clumps quickly collapse via gravitational instability and form planetesimals of characteristic size 100 km \citep{youdin05,johansen07,johansen15,simon16}.

Once planetesimals of that size are able to form, they need to grow in order to reach the mass of the giant planets' cores. This phase must happen before the gaseous disc dissipates, in order to allow giant planets to accrete the still-available gas after the core has grown. Considering only planetesimal accretion, it is challenging to grow giant planets' cores within the lifetime of the disc \citep{pollack96,rafikov04,levison10}. However, small solids observed in protoplanetary discs \citep{testi03,wilner05} have become a better candidate to be the main source of accretion for the growth of the giant planets' cores. 
The numerous embryos growing by pebble accretion gravitationally interact with one another so that the largest ones scatter the smaller ones out of the disc of pebbles, halting their growth and avoiding the production of too many Earth-mass objects \citep{levison15}. Therefore, the rapid and efficient pebble accretion significantly reduces the time scale of the growth of the core \citep{johansen10,ormel10,lambrechts12,ida16,johansen17} that becomes consistent with the lifetime of the gas-phase of the protoplanetary disc and forms only a few large cores rather than many smaller ones, consistently with what we observe in our solar system.

While the core is growing, it can become massive enough to retain an atmosphere in hydrostatic equilibrium because the heat generated by the accretion of pebbles and planetesimals prevents the contraction of the atmosphere \citep{lambrechts14,venturini17}. Once the accretion of pebbles onto the core stops, at the so-called `pebble isolation mass' \citep{lambrechts14b}, the atmosphere is no longer sustained by accretion energy and it contracts starting a phase of rapid gas accretion \citep{pollack96}. This process continues until the planet becomes massive enough to open a gap in the protoplanetary disc, thus slowing down the rate of gas supply, or until the protoplanetary disc photoevaporates. 

During the whole growth of the giant planet, interactions with the gaseous protoplanetary disc forces the protoplanet to migrate through the disc \citep{ward97}. Low mass planets are affected by the rapid Type-I migration, but once the planet becomes massive enough to open a gap in the disc, it migrates with a slower Type-II migration \citep{lin86}.

In this paper, we use giant planets' growth tracks of a solar system analogue, generated similarly to \citet{bitsch15b} (hereafter BLJ15). In BLJ15, the pebble flux, $\dot{\rm{M}}_{\rm{peb}}$, was overestimated and the correction highlighted the fact that the drift-limited pebble growth model in an evolving protoplanetary disc was an inadequate description \citep{bitsch18}. We here use the \citet{ida16} disc model with $\alpha=0.001$ (which reduces type-II migration compared to BLJ15) to generate the growth tracks of the solar system planets, which yields results very similar to the original growth tracks presented in BLJ15. We implemented them into the \textsc{Mercury} \textit{N}-body code \citep{chambers99} to analyse which consequences a large-scale migration, as the one predicted in BLJ15, could have on the minor bodies populations of our solar system. We will focus on the Jupiter Trojan asteroids and the Hilda asteroids that are the largest known populations in the Jovian mean motion resonances. The stability of these asteroid populations and their features are directly connected to the orbital configuration of the giant planets and because of this, they are sensitive to a possible migration. Therefore, the characteristics of the currently observed asteroids in resonance with Jupiter can contain information about the early evolution of the giant planets. 

The paper is organised as follows: in section \ref{sec:th} we summarise the characteristics of the Jupiter Trojan and Hilda minor body populations and the main hypothesis about their origin; in section \ref{sec:methods} we describe the methodology used in our simulations and the different scenarios we tested; in section \ref{sec:results} we present our results about the Jupiter Trojans and their asymmetry ratio, the Hilda group and the asteroid belt contamination; in section \ref{sec:nice} we explore, after the early migration, the effects of a rearrangement of the semimajor axes of the giant planets into their current ones. Finally, in section \ref{sec:conclusions} we summarise our results and their implications for planet formation and migration.
\section{Characteristics of the Trojan and Hilda populations}\label{sec:th}
\subsection{The Jupiter Trojan asteroids}

The Jupiter Trojan asteroids are bodies in the coorbital 1:1 resonance with Jupiter, librating around the two triangular equilibrium points in the Sun-Jupiter system, that is the Lagrangian points L$_4$ and L$_5$. The population that librates around L$_4$ represents the leading group, being located $60^\circ$ ahead of Jupiter along its orbit, called the `Greek camp'; the population that librates around L$_5$, located at about $60^\circ$ behind Jupiter along its orbit, is the trailing group, called the `Trojan camp'. Our two gas giants are close to their 5:2 resonance and studies about the long term dynamical stability of the Jupiter Trojan asteroids showed that their orbits are not indefinitely stable because of the gravitational interactions between Jupiter and Saturn, as a result of which they are slowly dispersing \citep{levison97}.

The number of detected Jupiter Trojans in the Minor Planet Center database (MPC hereafter), updated on the 30th of May 2018, is 7180\footnote{\url{https://minorplanetcenter.net/iau/lists/Trojans.html}}. The estimated total number of Trojans larger than 2 km in diameter is about $10^5$ asteroids \citep{nakamura2008b}. As regards the total mass, it is estimated to be $10^{-5}$ M$_\oplus$ \citep{vinogradova15}. We can summarise the main characteristics of the Jupiter Trojan asteroids as follows:

\begin{itemize}

\item The leading group is observed to be more populated than the trailing group. \citet{grav11}, with a sample detected by NEOWISE/WISE with sizes larger than 10 km, found a ratio of $1.4 \pm 0.2$ between the number of Trojans in L$_4$ and L$_5$. According to the Sloan Digital Sky Survey Moving Object Catalogue, in a kinematically selected sample of candidate Jovian Trojan asteroids with $H<13.8^m$ (approximately corresponding to 10 km diameter), there are $1.6 \pm 0.1$ more objects in the leading than in the trailing swarm \citep{szabo07}.  \citet{nakamura2008b}, using the observed sky number densities of L$_4$ and L$_5$ swarms with the Subaru telescope, estimated an asymmetry ratio of about $1.8 \pm 0.57$ for Trojans with sizes larger than $D>2$ km. Finally, \citet{vinogradova15} estimated an asymmetry ratio of about 2.0 for Trojans with absolute magnitude $H<19^m$ (sizes larger than $D>1$ km). We also analysed the Trojans in the MPC\footnote{{\url{https://minorplanetcenter.net/iau/lists/t_jupitertrojans.html}}} with H<12 (complete sample according to \citet{szabo07}). In the list there are 330 L$_4$ objects and 229 L$_5$ objects. As in \citet{nesvorny13}, we excluded 9 Eurybates family members from L$_4$ finding an asymmetry ratio of about $1.4 \pm 0.1$, consistent with the \citet{grav11} constraint.

\item Jupiter Trojans are characterised by high inclinations, up to 40$^\circ$, and their distribution differs between the L$_4$ and L$_5$ swarms. The L$_5$ population shows a significantly wider slope distribution with a plateau in the range from $5^\circ$ to $17^\circ$ and a weak maximum at $27^\circ$. The distribution of Trojans in L$_4$ shows a sharp maximum at $7^\circ$, after which the number of Trojans exponentially decreases \citep{slyusarev14}. 

\item The Trojan population is more homogeneous than the asteroid belt population. They are very dark, mainly P and D type asteroids \citep{demeo14}, and they have a featureless, red-sloped spectra at visible and near-infrared wavelengths \citep{barucci02}. Trojan asteroids with high inclination tend to be redder in a similar way for both swarms, but there is no correlation between the colour of the asteroid and the object's size \citep{szabo07}. Moreover, the optical colour distributions of the Jovian and Neptunian Trojans are indistinguishable from each other, but they are statistically different from the Kuiper belt populations \citep{jewitt18}.

\end{itemize}

Regarding the origin of the Jupiter Trojans, it is still a matter of debate. The main hypotheses are:
\begin{enumerate}[(i)] 
\item Collisions of primordial planetesimals that can inject fragments into Trojans orbits \citep{shoemaker89}.
\item Drift into the Trojan regions due to the action of a dissipative forces like gas drag or the Yarkovsky effect \citep{yoder79,peale93,kary95}.
\item An in situ capture of the Trojans from the feeding zone by a growing proto-Jupiter \citep{shoemaker89,pollack96,marzari98a,marzari98b,fleming00}. 
\item A capture of bodies originally formed in the outer solar system and injected in the inner regions during a possible late instability of the giant planets. The capture can happen during the crossing of the 2:1 resonance between Jupiter and Saturn, while they are undergoing a divergent migration, in a process called `chaotic capture' \citep{morbidelli05}.
\item A `jump capture' \citep{nesvorny13}. Trojans are captured after Jupiter suffered a close encounter with an ice giant. As a result, the semimajor axis of Jupiter `jumps' and radially displaces the L$_4$ and L$_5$ regions, losing the existing Trojans and capturing new bodies with semimajor axis similar to the new position of Jupiter.
\end{enumerate}
Hypotheses (i), (ii), (iii) cannot reproduce the high inclinations, the asymmetry and other observational constraints \citep{marzari02}. That is why the last two hypotheses are the most accepted ones today: hypothesis (iv) offers an explanation for the inclination distribution and for the outer solar system origin of the Trojans, though the capture probability is very low \citep{lykawka10} and the model cannot explain the asymmetry ratio between the two Trojan swarms; hypothesis (v) reproduces the orbital distribution of the Trojans and it is also potentially capable of explaining the asymmetry ratio of the Trojans in case the extra ice giant involved in the planet-planet scattering with Jupiter traverses one of the Lagrange swarms and scatters captured bodies, depleting the swarm. In this scenario, even if the extra ice giant traverses the correct swarm, the asymmetry found in \citet{nesvorny13} cannot rule out symmetric swarms within $1\sigma$. Also, the low capture probability could still be considered as a weakness. Our model, for the first time, supplies a natural explanation for the Trojan asymmetry and their photometric colour.


\subsection{The Hilda asteroids}

The Hilda asteroids are in the 3:2 orbital mean motion resonance with Jupiter. They move along orbits with a semimajor axis between 3.7 au and 4.2 au with eccentricities up to 0.3 and inclinations up to $20^\circ$. Currently (May 2018), the MPC database contains 4008 objects with an Hilda asteroid orbit type\footnote{\url{https://minorplanetcenter.net/db_search/show_by_orbit_type?utf8=\%E2\%9C\%93&orbit_type=8}}. The total number of the Hilda population larger than 2 km in diameter is estimated to be $\sim1 \times 10^4$ based on the size distribution \citep{terai18}.
Regarding their stability, Hilda asteroids are in a mean motion resonance and even if many of their orbits are definitely chaotic because of this, the resulting escape times does not conflict with their presence today \citep{franklin93}.

\begin{itemize}
\item As the Jupiter Trojan asteroids, Hildas' surface photometric colours often correspond to the low-albedo D-type and P-type. In contrast, the asteroid belt is dominated by C-type asteroids, but a small fraction of D/P types is also present in the belt \citep{demeo14}. 
\item An analysis of the near-infrared spectra of 25 Hilda asteroid reveals that red Hildas and Trojans have nearly identical spectra, whereas the less-red Hilda asteroids are significantly redder than less-red Trojans in the near-infrared \citep{wong17}. The authors suggested that the discrepancy between less-red Hilda and Trojan spectra derives from the present-day temperature gradient between the Hilda and Trojan regions, while red spectral types are not sensitive to this gradient.
\item The size distributions of Hilda and Jupiter Trojan asteroids are very similar to each other within the range of diameter $2<D<10$ km, while these distributions are distinguishable from that of main belt asteroids, suggesting a common origin and a different formation environment from main-belt asteroids \citep{terai18}.
\end{itemize}

One of the hypothesis for the origin of the Hilda asteroids is, as for the Trojan asteroids, the `chaotic capture' of bodies from the external region of the solar system after a late instability of the giant planets \citep{morbidelli05}. Simulations have shown that the primordial Hilda population was destabilised and lost during the late instability and then replaced with planetesimals scattered inward from the external region of the solar system \citep{gomes05,roig15}.
It is also possible that, in the scenario where there were originally five giant planets \citep{nesvorny11}, planetesimals from the outer solar system are captured in the inner, central and outer asteroid belt, as well as Cybele, and the resonant Hilda and Thule regions \citep{vokrouhlicky16}. 
Another hypothesis involves a capture of field asteroids if Jupiter has migrated sunward by about 0.45 au \citep{franklin04}. The authors pointed out that a migration of $\Delta a > 0.45$ leads to a poor match of the average eccentricity of the Hildas and an high percentage of Hilda asteroids with eccentricity $e> 0.2$ compared to the observed one. However, this study is confined to Jupiter and Saturn in their current ratio of periods or in their 5:2 resonance and with their current masses. No gas drag to mimic the presence of the protoplanetary disc is taken into account, as well as no primordial Hilda population and no possible rearrangement of the giant planets after the gaseous disc disperses, which we all include here.


\section{Methods}\label{sec:methods}

In our simulations, we used a parallelised version of the \textsc{Mercury} \textit{N}-body code \citep{chambers99} and we selected its hybrid symplectic integrator because it is faster than conventional N-body algorithms by about one order of magnitude \citep{wisdom91} and it is particularly suitable for our simulations that involve timescales of the order of $10^6 - 10^9$ years. We used a time step of 50 days which is about 1/20 of the orbital period of particle orbiting at 2 au \citep{duncan98}. Even if we populate our solar system from 4 au and beyond, with this time step we also correctly reproduce the dynamical evolution of particles that could end up in the asteroid belt region. We modified the code so that the giant planets grow and migrate following the growth tracks generated similarly to BLJ15, as will be discussed in section \ref{sec:growthtracks}. Moreover, our version of the code includes aerodynamic gas drag effects on the small particles and tidal gas drag effects on the planetary cores (see appendixes \ref{sec:agd} and \ref{sec:tgd}, respectively) to mimic the presence of a gaseous protoplanetary disc with a lifetime of 3 Myr.

\subsection{Migration model for the giant planets}\label{sec:growthtracks}

As mentioned in section \ref{sec:intro}, we generated growth tracks similar to the ones in BLJ15 where the giant planets of our solar system grow and migrate through the disc ending up in a more compact final configuration, as predicted by many hydrodynamical and N-body simulations \citep{masset01,kley04,morbidelli07,pierens08}. 
The protoplanetary disc model used for the growth tracks, for the aerodynamic gas drag disc parameters (appendix \ref{sec:agd}) and for the tidal gas drag (appendix \ref{sec:tgd}) disc parameters, is the \citet{ida16} disc model. This empirical disc model includes both viscous heating and stellar irradiation assuming a steady accretion disc with a constant $\alpha$-viscosity parameter (that we set to 0.001) and analytical formulas for the pebble accretion rate onto planetary embryos. As in BLJ15, we considered the evolution of the disc in time so the accretion rate $\dot{M}$ is not constant, but changes following the empirical law in \citet{hartmann98}:
\begin{equation}
\log{\left(\frac{\dot{M}}{M_\odot/\rm{yr}}\right)}=-8.00 -1.40 \log{\left(\frac{t}{10^6 \rm{yr}}\right)}
\end{equation}
Finally, for the gas accretion and planetary migration, we followed the prescriptions in BLJ15.

\begin{figure}
\begin{center}
\includegraphics[width=\hsize]{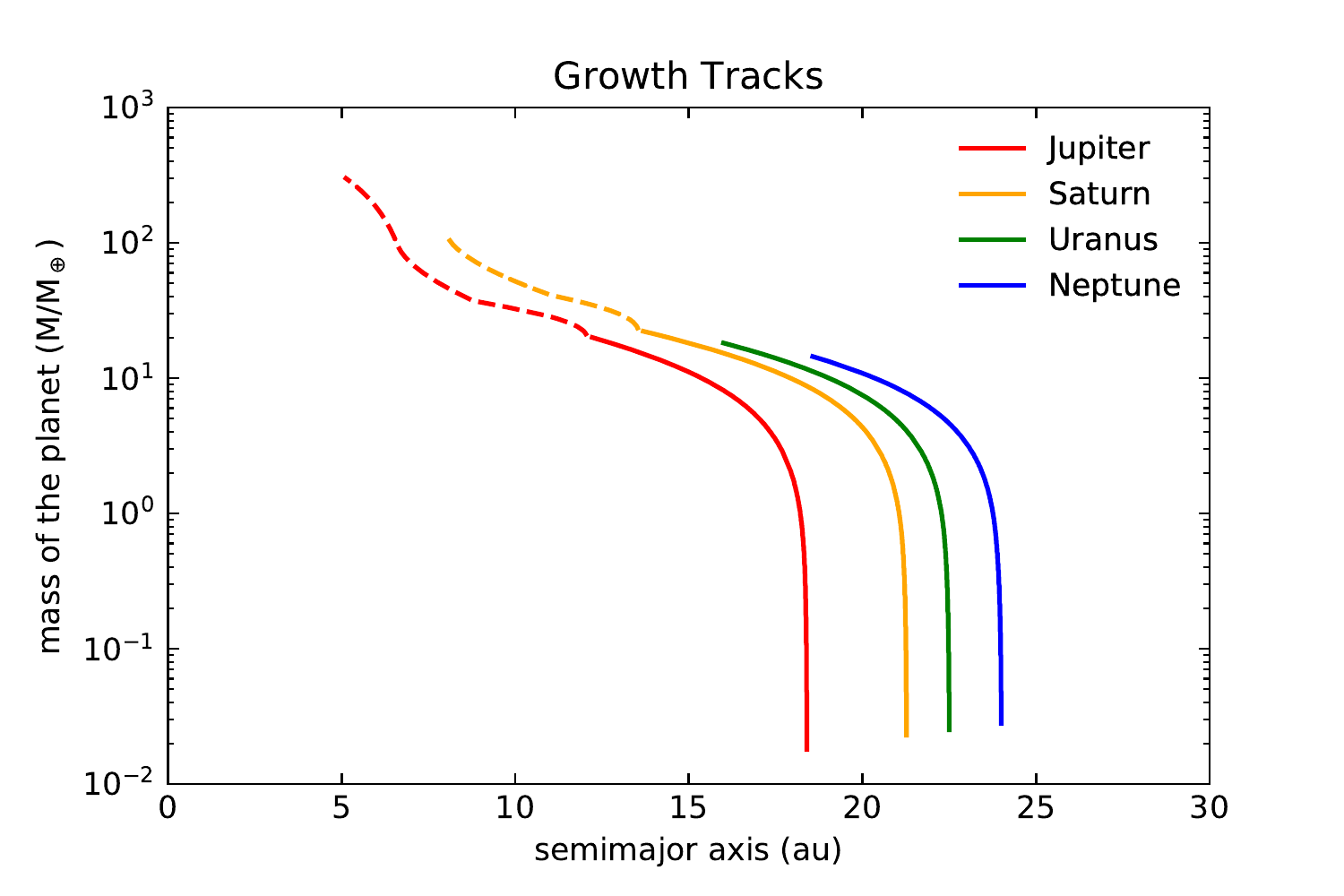}
\caption[]{The plot shows growth tracks for our nominal model. Jupiter (red line), Saturn (orange line), Uranus (green line) and Neptune (blue line) grow and migrate through the disc ending up in a more compact configuration compared to the current one. The solid lines indicate the core accretion phase, the dashed lines indicate the gas accretion phase.}
\label{fig:gt}
\end{center}
\end{figure}

Growth tracks of the giant planets for our nominal model are shown in Figure \ref{fig:gt}. Giant planet seeds, with masses of the order of  $10^{-2}$ M$_\oplus$, are implanted in the disc at about 18 au, 21 au, 23 au, 24 au, in order to get the current masses and a more compact orbital configuration at the end of the migration similarly to the initial conditions in \citet{malhotra95,morbidelli10}. Jupiter, Saturn, Uranus and Neptune start to grow and migrate at $t\sim2.31$ Myr, $t\sim2.56$ Myr, $t\sim2.70$ Myr and $t\sim2.70$ Myr, respectively, embedded in the gaseous protoplanetary disc that photoevaporates at $t=3$ Myr, that is when the accretion rate becomes lower than $\dot{M} \le 2 \times 10^{-9} \text{M}_\odot \text{yr}^{-1}$. The giant planet cores initially grow with relatively little migration, accreting pebbles (solid line in the Figure \ref{fig:gt}). While growing, the protoplanets become massive enough to be affected by Type I migration. When the gravity of the core is able to create a pressure bump that traps particles outside the accretion radius, the solid accretion halts and the core reaches its so-called `pebble isolation mass' and it starts to accrete gas (dashed line in the figure \ref{fig:gt}). If the mass of the envelope becomes larger than the core mass, the planet undergoes a runaway accretion of gas and, in case it becomes massive enough to open a gap in the protoplanetary disc, it experiences a slower Type-II migration. Eventually, when the disc dissipates at $t=3$ Myr, the four giant planets have reached masses similar to the current ones and they are orbiting within 5 and 20 au from the central star.

\begin{figure}
\begin{center}
\includegraphics[width=\hsize]{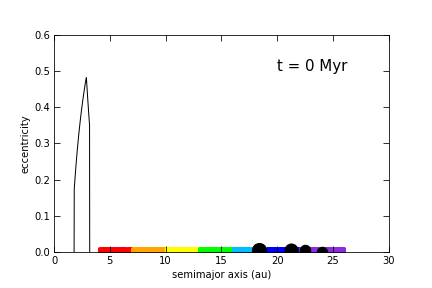}
\caption[]{Initial configuration of the solar system in the nominal model. The plot shows colour-coded small bodies, from 4 to 26 au, starting with eccentricities in the $[0:0.01]$ range. We used different colours for the small particles to easily track them during the simulations. The black filled dots at 18 au (Jupiter), 21 au (Saturn), 22 au (Uranus) and 24 au (Neptune) are the giant planets seeds. The black line roughly delimits the actual main belt region, from 2 au to 3.4 au, with a cut for perihelion distances smaller than 1.7 (particles strongly affected by the presence of Mars) and a cut for aphelion distances greater than 4.5 (particles strongly affected by the presence of Jupiter).}
\label{fig:ploti0o}
\end{center}
\end{figure}

We populate annular regions of width $0.5$ au with $1000$ massless particles each. The same amount of particles in each annular region means a surface density proportional to $r^{-1}$ for this component consistent with the $\Sigma_g$ (column density of the gas) slope in the outer disc in the \citet{ida16} disc model. We performed a first set of 10 simulations involving 44000 small bodies from just outside the main asteroid belt to the region affected by the orbit of the last protoplanet, that are, in our nominal model, 4 and 26 au, respectively. In each simulation the growth tracks are kept the same, but we randomised the small body eccentricities in the interval $[0,0.01]$, the inclinations in the interval $[0^\circ,0.01^\circ]$ and the semimajor axes in every $\Delta a=0.5$ au annular region, in order to get more robust results. As shown in Figure \ref{fig:ploti0o} planetary cores and small particles started with low eccentricities and low inclinations. The region between 0 and 4 au is left completely empty, that is no small body populates the terrestrial planet and the main asteroid belt regions and the black line inside this inner region, roughly delimits the actual main belt region. It is important to notice that, even if we have chosen not to populate the asteroid belt for computational time reasons, we do not consider this region empty. Indeed, the main belt is populated with `virtual' stony (silicaceous) objects and carbonaceous objects (S-type and C-type asteroids, respectively) and, taking into account the Minimum Mass Solar Nebula \citep{weidenschilling77,hayashi81}, the total mass of the belt is about 1 M$_\oplus$.

In the \textsc{Mercury} \textit{N}-body code, the planets are treated as massive bodies, so they perturb and interact with all the other bodies during the integration. The other particles, called small bodies, are set as massless, so they are perturbed by the massive bodies but cannot affect each other nor the massive bodies. 
The growing protoplanets, in our modified version of the \textsc{Mercury} \textit{N}-body code, are affected by the tidal gas drag, as described in appendix \ref{sec:tgd}. The small bodies are affected by the the aerodynamic gas drag until the gaseous disc photoevaporates at $t=3$ Myr. Since they are massless, we assign them a radius $r_{\rm{p}}=50$ km 
and a density $\rho_{\rm{p}}=1.0$ g/cm$^3$ when computing the effect of the aerodynamic gas drag, as explained in Appendix \ref{sec:agd}.

\subsection{In situ growth simulations}\label{sec:insitusim}

As a comparison to the nominal model we want to test, we also simulate the in situ growth of the gas giants. In this particular case, Jupiter and Saturn seeds are implanted at 5.4 au and 8.6 au, respectively (Table \ref{table:difrates}), in order to approximately match the same final configuration of our nominal model, so we can compare the results. We initially tested the inclusion of the ice giants in the in situ simulations and concluded that their contribution to Trojan and Hilda asteroids is negligible, therefore we opted just for the presence of the gas giants. The growth rate is kept the same as in the nominal model: Jupiter and Saturn start to grow at $t\sim2.31$ Myr and $t\sim2.56$ Myr, respectively, and stop at $t=3$ Myr when the disc dissipates. They are not forced to migrate. Small bodies starting eccentricities and inclinations are the same as in the nominal model, randomised in each of the 10 runs. Aerodynamic and tidal gas drag are taken into account as well.

\subsection{Different migration rate simulations}\label{sec:diffrates}

To further explore the effects of planetary migration, we tested different migration rates for the growth tracks. For this subset of simulations we took into account only Jupiter and Saturn. The two gas giants start to migrate at the same time as the growth tracks in the nominal model and stop at $t=3$ Myr as well, therefore the growth rates of the planets are kept the same. 
\begin{figure}
\begin{center}
\includegraphics[width=\hsize]{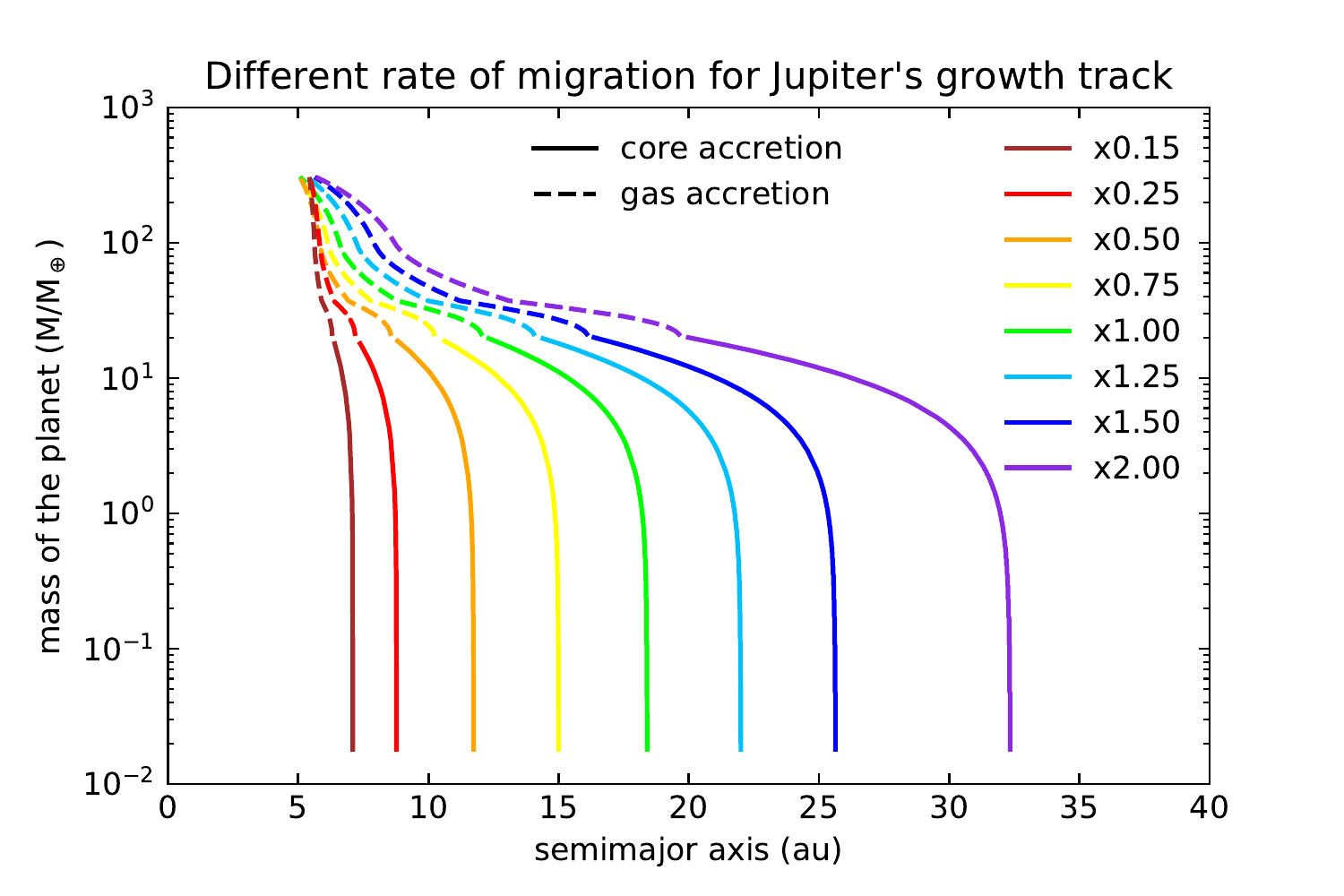}
\caption[]{Different rates of migration for Jupiter growth tracks compared to the nominal migration rate (green line). The brown, red, orange and yellow curves are associated with slower migration rates: 0.15, 0.25, 0.50 and 0.75 times the nominal rate of migration, respectively. The blue, dark blue and violet curves are associated with faster migration rates: 1.25, 1.50 and 2.00 times the nominal migration rate, respectively. The solid lines indicate the core accretion phase and the dashed lines indicate the gas accretion phase. Due to interactions between the gas giants during the migration, Jupiter does not exactly end up with the same semimajor axis for each migration rate.}
\label{fig:rates}
\end{center}
\end{figure}
Faster migration rates mean implanting Jupiter and Saturn's seeds further away in the outer solar system compared to the nominal model. Slow migration rates mean implanting the two giant planets' seeds closer to the central star than in the nominal model. Hence, these growth tracks are just a contraction or expansion of the nominal model growth tracks and they represent just a simplification. A plot of the different growth tracks with different migration rates for Jupiter is shown in Figure \ref{fig:rates}. We tested the following slower migration rates: 0.15, 0.25, 0.50 and 0.75 times the original rate of migration and the following faster migration rates: 1.25, 1.50 and 2.00 times the original migration rate. 
\begin{table}
\caption{Initial semimajor axis of Jupiter and Saturn' seeds for the different migration rate.}             
\label{table:difrates}      
\centering                          
\begin{tabular}{clc lc l  }        
\hline\hline               
Migration rate & \multicolumn{2}{c}{Starting semimajor axis} \\
(times the nominal & \multicolumn{2}{c}{of the giant planets}\\
 \cline{2-3}
migration rate)& $a_{\rm{J}}$  & $a\rm{_S}$ \\    
\hline                        
    in situ  & $5.4$  & $8.6$ \\
    $\times$0.15  & $7.1$ & $10.45$ \\
    $\times$0.25  & $8.8$ & $11.9$ \\
    $\times$0.50  & $11.7$ & $14.6$ \\
    $\times$0.75   & $15.0$ &  $18.0$  \\
    $\times$1.00& $18.4$ &  $21.3$ \\
    $\times$1.25  & $22.2$ &  $25.5$ \\      
    $\times$1.50  & $25.6$ & $28.9$ \\
    $\times$2.00  & $32.3$ &  $35.5$  \\ 
\hline                                  
\end{tabular}
\end{table}
Initial semimajor axes of Jupiter and Saturn's seeds for different migration rates are listed in Table \ref{table:difrates}.
For each migration rate, we ran ten simulations randomising the initial semimajor axes in every $\Delta a$, the eccentricities and the inclinations of the small bodies, as we did in the nominal model simulations and the in situ model simulations.


\section{Results}\label{sec:results}

\begin{figure*}
\begin{center}
\includegraphics[width=8.3 cm]{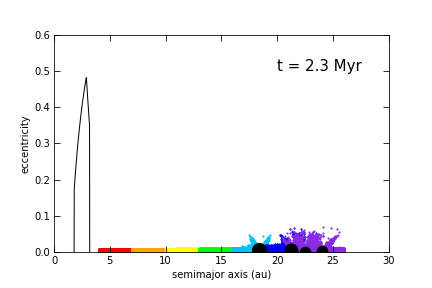}
\includegraphics[width=8.3 cm]{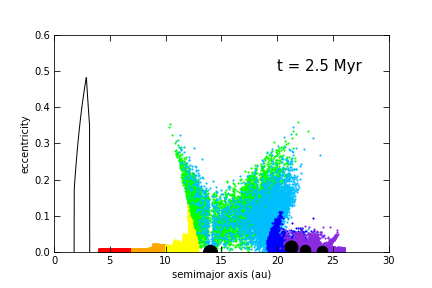}
\includegraphics[width=8.3 cm]{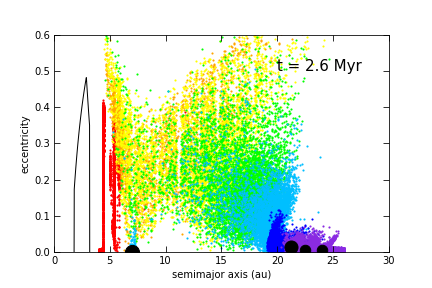}
\includegraphics[width=8.3 cm]{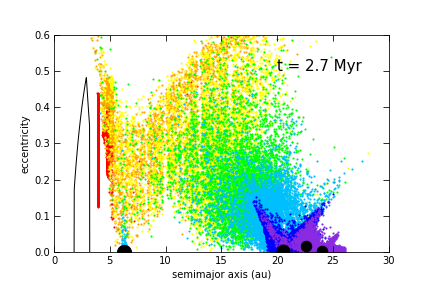}
\includegraphics[width=8.3 cm]{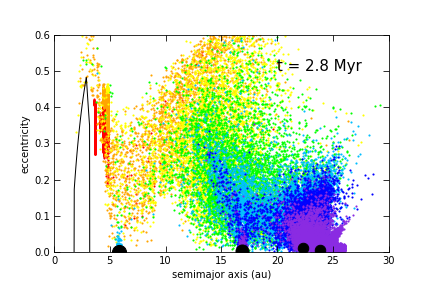}
\includegraphics[width=8.3 cm]{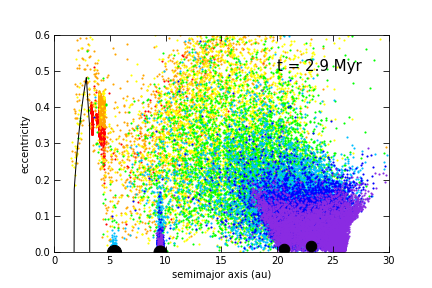}
\includegraphics[width=8.3 cm]{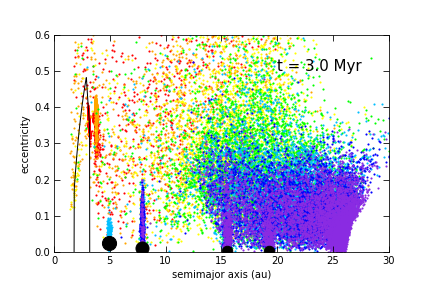}
\includegraphics[width=8.3 cm]{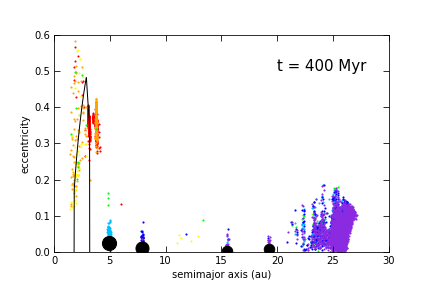}
\caption[]{The figure shows eight time-snapshots of eccentricity versus semimajor axis of one of the nominal model simulations. The black line roughly delimits the main belt region. At $t=2.31$ Myr, $t=2.56$ Myr, $t=2.70$ Myr and $t=2.70$ Myr, Jupiter, Saturn, Uranus and Neptune (black filled dots) start to grow and to migrate inward and stop at $t=3$ Myr. A fraction of bodies that were initially placed between 4 and 16 au, are injected in the asteroid belt or possess enough eccentricity to cross the terrestrial planet region. Trojan asteroids are captured in the feeding zone of the planets' cores and carried along during the migration. Each of the four planets captured Trojans, but, after $t=400$ Myr, Saturn's, Uranus' and Neptune's are drastically reduced because of planetary interactions. Particles clump in the resonance 3:2 (the Hilda asteroids at roughly 4 au) and other populated resonances are shown in Figure \ref{fig:res}. A small population of centaurs orbits between Jupiter and Neptune.}
\label{fig:plote100}
\end{center}
\end{figure*}

\begin{figure*}
\begin{center}
\includegraphics[width=8.3 cm]{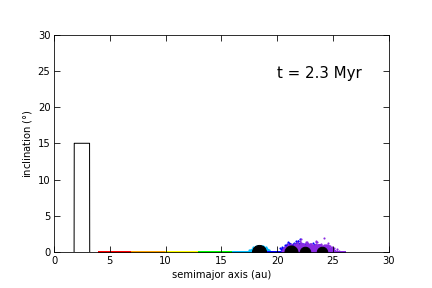}
\includegraphics[width=8.3 cm]{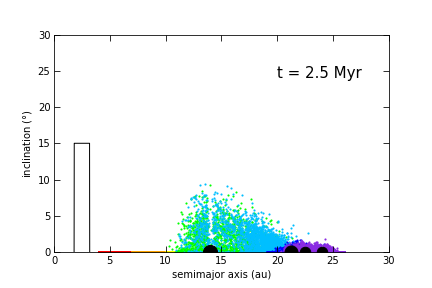}
\includegraphics[width=8.3 cm]{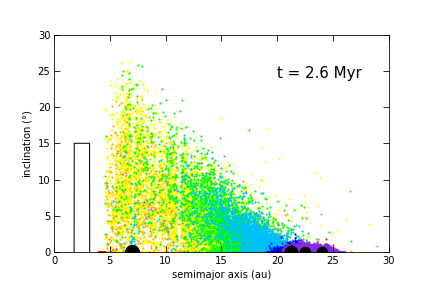}
\includegraphics[width=8.3 cm]{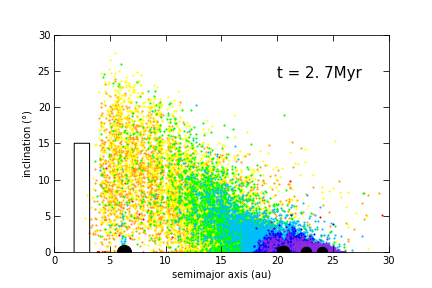}
\includegraphics[width=8.3 cm]{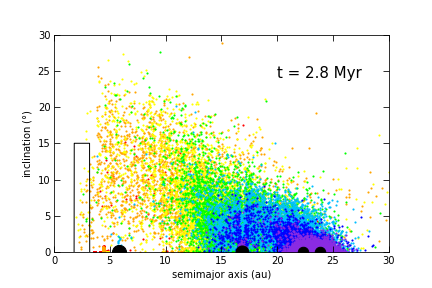}
\includegraphics[width=8.3 cm]{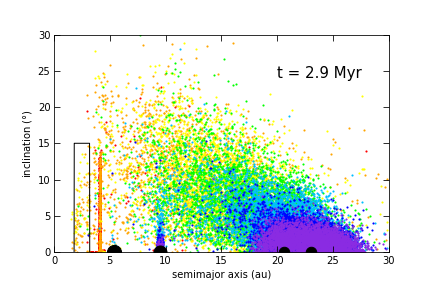}
\includegraphics[width=8.3 cm]{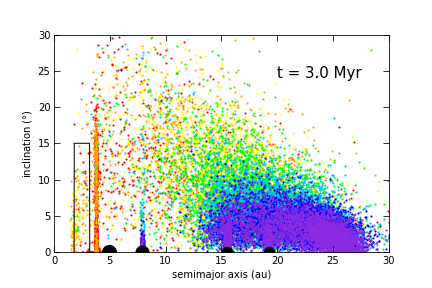}
\includegraphics[width=8.3 cm]{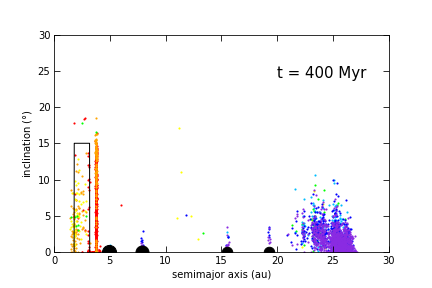}
\caption[]{As Figure \ref{fig:plote100}, eight time-snapshots of inclination versus semimajor axis of one of the nominal model simulations are shown. The black line roughly delimits the main belt region. At $t=2.31$ Myr, $t=2.56$ Myr, $t=2.70$ Myr and $t=2.70$ Myr, Jupiter, Saturn, Uranus and Neptune (black filled dots) start to grow and to migrate inward and stop at $t=3$ Myr. A fraction of bodies that were initially placed between 4 and 16 au, are injected in the asteroid belt region. Trojan asteroids are captured in the feeding zone of the planets' cores and carried along during the migration. As can be noted, particles grouped in the resonance 3:2 (the Hilda asteroids at roughly 4 au), gain their inclination because of Saturn migrating towards Jupiter. A small population of centaurs orbits between Jupiter and Neptune.}
\label{fig:ploti100}
\end{center}
\end{figure*}

In our nominal model, Jupiter, Saturn, Uranus and Neptune start to grow and migrate at $t=2.31$ Myr, $t=2.56$ Myr, $t=2.70$ Myr and $t=2.70$ Myr, respectively. At $t=3$ Myr, the planets are fully grown and have migrated to their final positions. Figures \ref{fig:plote100} and \ref{fig:ploti100} show time-snapshots of their eccentricity versus semimajor axis and inclination versus semimajor axis, respectively, of the same simulation. The black line roughly delimits the actual main belt region, but shifted 0.2 au inward, since Jupiter ended up at about 5.0 au. The plots clearly show that the large scale migration deeply shapes the solar system and that the small particles are strongly affected by the passage of the giant planets. The main consequences of this early migration are:

\begin{figure}
\begin{center}
\includegraphics[width=\hsize]{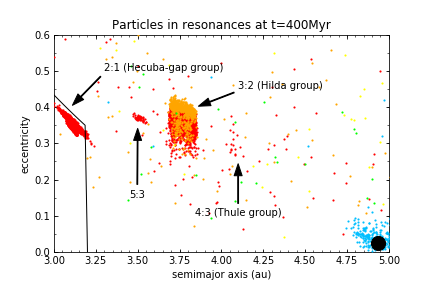}
\caption[]{Zoom in of the region between 3 and 5 au in figure \ref{fig:plote100}. It is possible to distinguish different resonant populations. From the left to the right in the plot: the Hecuba-gap group in the 2:1 mean motion resonance with Jupiter (3-3.25 au); a population in the second order 5:3 resonance with Jupiter (3.5 au); the Hilda asteroids in the 3:2 mean motion resonance with Jupiter (3.75 au); the Thule group in the 4:3 resonance (4.1 au). The black line roughly delimits the main belt region. We note that in this nominal case, due to interactions with the other planets, the final semimajor axis of Jupiter is about 5 au and the location of the resonances are slightly shifted inward.}
\label{fig:res}
\end{center}
\end{figure}

\begin{itemize}
\item A fraction of bodies that were initially placed between 4 and 16 au, are injected in the asteroid belt or possess enough eccentricity to cross the terrestrial planet region, contaminating the inner solar system;
\item All four giant planets captured Trojans from the feeding zone of their cores and carried them along during the migration; 
\item Particles are captured in the first order resonances 2:1, 3:2 and 4:3 with Jupiter and in the second order resonance 5:3 as showed in Figure \ref{fig:res}; 
\item A centaur population orbits between Saturn and Uranus. 
\end{itemize}


\subsection{Characterisation of the Jupiter Trojan asteroids}

During planetary growth and migration, each giant planet captured and preserved a population of asteroids in their 1:1 coorbital resonance, even if not all of them showed a long term stability as the Jupiter's Trojans. Indeed, as shown in Figures \ref{fig:plote100} and \ref{fig:ploti100}, after $t=400$ Myr Saturn, Uranus and Neptune Trojan asteroids are drastically reduced in number because of planetary interactions.

\subsubsection{Trojan asymmetry}\label{sec:ta}

In our nominal model, the Jupiter Trojan asteroids found in all the 10 simulations are always characterised by a certain ratio of asymmetry between the number of Trojan asteroids in the leading group and the trailing group, with the L$_4$ swarm always more populated than the L$_5$ swarm. 
\begin{table}
\caption{Mean quantities of the Jupiter Trojans in the nominal model.}             
\label{table:trojan_table}     
\centering                          
\begin{tabular}{r c c c }        
\hline\hline                 
 Time (Myr) & $\rm{N_{tot}}$  & $\rm{N_{L_{4}}/N_{L_{5}}}$ & \\    
\hline                      
    $5$  & $240\pm 9$  & $ 1.80\pm 0.26 $ & \\
    $10$  & $231\pm 8$ & $ 1.85\pm 0.28$ & \\
    $50$  & $222\pm 7$ &  $ 1.88\pm 0.28$ &  \\
    $100$  & $218\pm 7$ &  $ 1.90\pm 0.27$ & \\
    $200$  & $214\pm 8$ &  $ 1.92\pm 0.29$ & \\      
\hline                                   
\end{tabular}
\end{table}
In Table \ref{table:trojan_table}, we report the mean total number of Trojans and the asymmetry ratio between the number of particles in the L$_4$ and L$_5$ swarms at different times. We computed the arithmetic mean of the values found in the 10 simulations and the uncertainty is represented by the unbiased standard deviation. For the asymmetry ratio error, a propagation of the uncertainty is applied. 

\begin{figure}
\begin{center}
\includegraphics[width=\hsize]{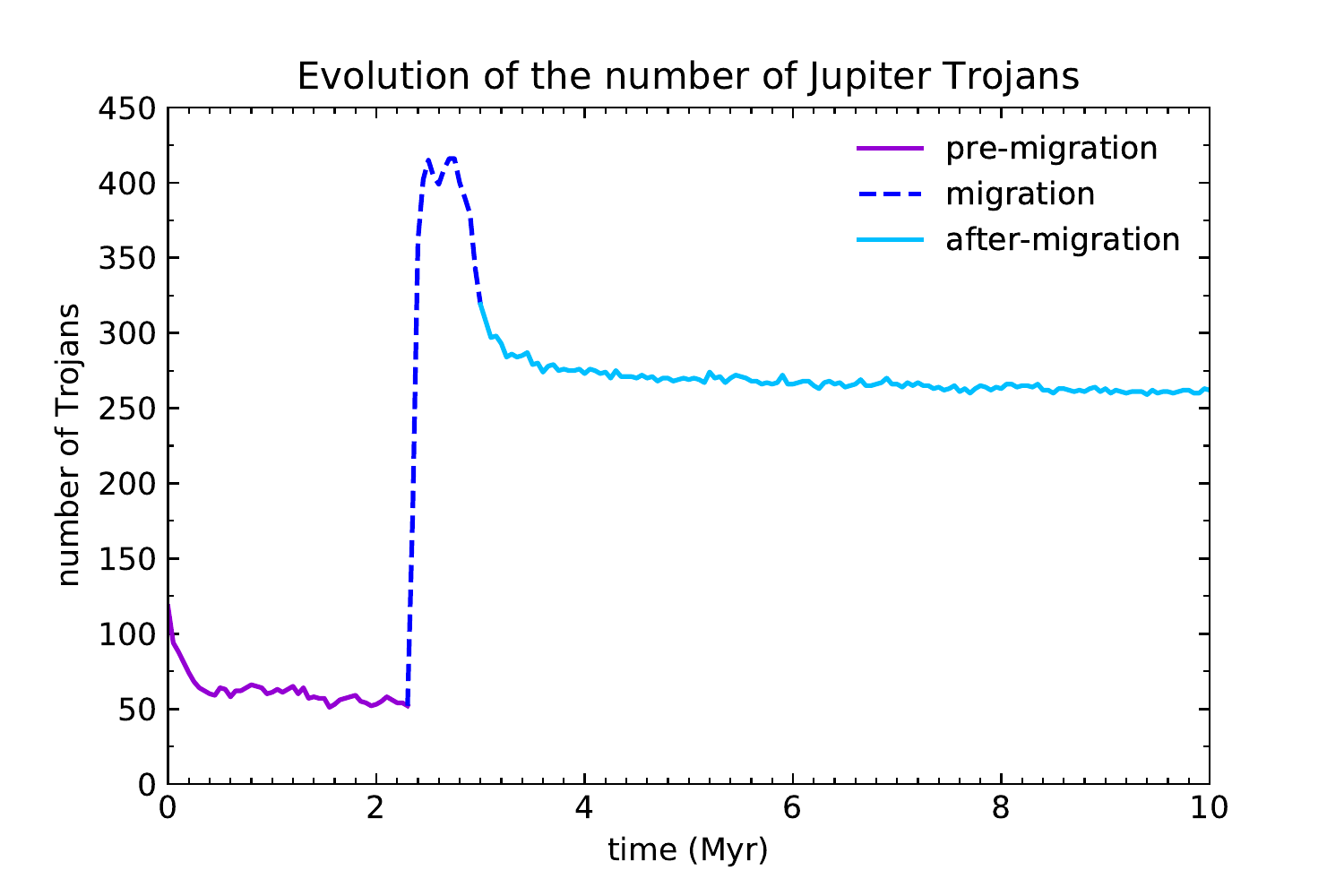}
\includegraphics[width=\hsize]{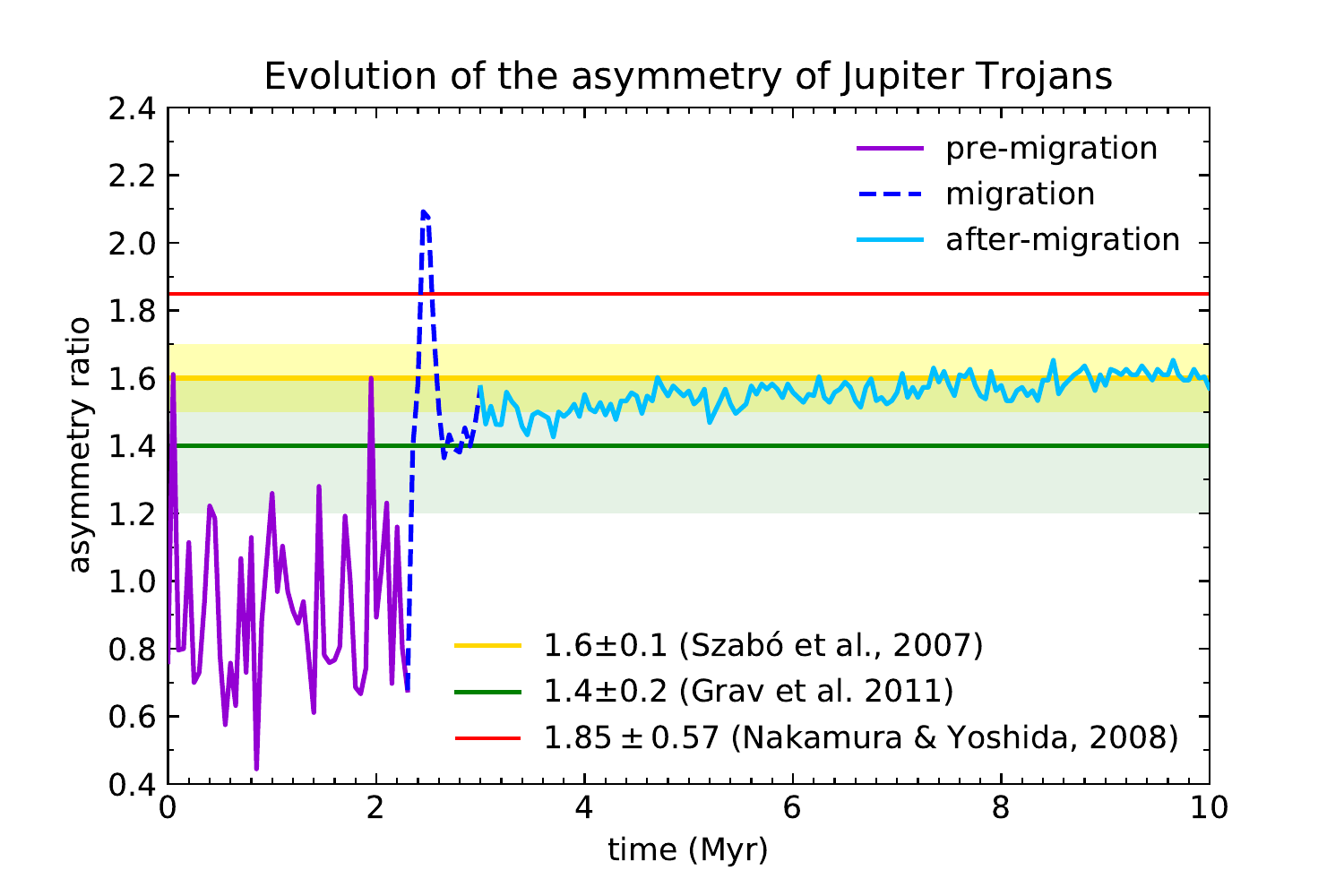}
\caption[]{Evolution of the number of Trojan asteroids (top plot) and the asymmetry ratio between the L$_4$ and L$_5$ swarms (bottom plot). The migration of Jupiter starts at $t\sim2.3$ Myr and stops at $t=3$ Myr, when the protoplanetary disc photoevaporates. The green shaded region represents the observed values of the Trojan asymmetry found by \citet{grav11}, the yellow shaded region represents the observed values of the Trojan asymmetry found by \citet{szabo07} and the red line represents the asymmetry ratio estimated in \citet{nakamura2008b}. In this latter case the shaded region has not been overplotted to keep the figure readable, but the uncertainty is indicated in the legend.}
\label{fig:asymmetry}
\end{center}
\end{figure}

Figure \ref{fig:asymmetry} shows an example of the evolution of the number of the Trojans (top plot) and the asymmetry ratio (bottom plot) in the first 10 Myr of one of the 10 simulations of our nominal model. We consider as Trojans all those particles that have aphelion and perihelion within $\pm 2.5$ Jupiter Hill radii from the planet semimajor axis. This way we include almost all the main coorbital resonances: tadpole orbits (Trojans, that librate around $\pm 60^\circ$), horseshoe orbits (objects that librate around 180$^\circ$ encompassing L$_4$ and L$_5$) and, occasionally, quasi-satellite orbits (asteroids that librate around 0$^\circ$, i.e. around Jupiter's position), excluding interlopers with high eccentricity.
Following the growth tracks, Jupiter undergoes the migration and grows at the same time. While the migration destabilises the Trojans, the rapid growth expands the stability regions around L$_4$ and L$_5$ and reduces the libration angles of the Trojans as found in \citet{marzari98a} and \citet{fleming00}, so initial horseshoe orbits become L$_4$ or L$_5$ tadpole orbits. Just after $t=2.31$ Myr the destabilising effect of the migration is completely suppressed by the growth of the planet that permits to capture and keep stable almost 7 times more Trojans than the ones present in the pre-migration phase. However, Jupiter stops trapping Trojans while migrating inward likely because the interior mean motion resonances pump up the eccentricities of possible Trojans above the Hill eccentricity, leading to a capture inefficiency. At $2.7$ Myr, the number of Trojans declines towards a total number that is 5 times more than the pre-migration one. At $t=3.2$ Myr the number of Trojans has eventually stabilised against the effect of the migration and growth, but keeps slowly decaying due to fact that Jupiter and Saturn are in their 1:2 resonance and it is an unstable configuration particularly for Saturn Trojans, but also for Jupiter ones \citep{gomes98}. Figure \ref{fig:asymmetry}, bottom plot, also shows that the asymmetry ratio slightly increases after the dispersal of the gas, this means that the 1:2 resonance between the giant planets configuration preferentially destabilises the L$_5$ swarm (see also Table \ref{table:trojan_table}).

\begin{table}
\caption{Mean quantities of the Trojans at $t=5$ Myr, with different migration rate of the giant planets}             
\label{table:rates}      
\centering                        
\begin{tabular}{l c c  }       
\hline\hline                 
 Migration rate & $\rm{N_{tot}}$  & $\rm{N_{L_{4}}/N_{L_{5}}}$ \\    
\hline                       
    in situ  & $51\pm 8$  & $0.94 \pm 0.35$ \\
    $\times$0.15   & $82\pm 9$ & $1.23 \pm 0.40$ \\
    $\times$0.25   & $106\pm 11$ & $1.60 \pm 0.44$ \\
    $\times$0.50  & $165\pm 11$ & $1.63 \pm 0.29$ \\
    $\times$0.75 & $216\pm 9$ &  $1.73 \pm 0.29$  \\
    $\times$1.00& $240\pm 9$ &  $1.80 \pm 0.26$ \\
    $\times$1.25  & $289\pm 12$ &  $1.44 \pm 0.15$ \\      
    $\times$1.50  & $361\pm 13$ & $1.63 \pm 0.21$ \\
    $\times$2.00 & $419\pm 21$ &  $1.77 \pm 0.27$  \\ 
    \hline
    \multicolumn{3}{c}{observed asymmetry ratios}\\
    \hline
    \multicolumn{2}{c}{\citet{grav11}}&$1.4 \pm 0.2$\\
    \multicolumn{2}{c}{\citet{szabo07}}&$1.6 \pm 0.1$\\
    \hline
    \multicolumn{3}{c}{estimated asymmetry ratio}\\
    \hline
    \multicolumn{2}{c}{\citet{nakamura2008b}}&$1.85 \pm 0.57$\\
    \hline                                   
\end{tabular}
\end{table}

As introduced in section \ref{sec:insitusim}, we ran additional simulations with an in situ growth of the giant planets. The resulting mean total number of Trojan captured and their asymmetry ratio are listed in Table \ref{table:rates}. The capture of the Trojans in the in situ model is less efficient by roughly a factor 4 and, more important, the in situ growth produced Trojans with no asymmetry between the number of bodies populating the two swarms. Moreover, roughly half of the simulations ended up with an asymmetry ratio opposite to the observed one, that is with the L$_4$ swarm less populated than the L$_5$ swarm.

\begin{figure}
\begin{center}
\includegraphics[width=\hsize]{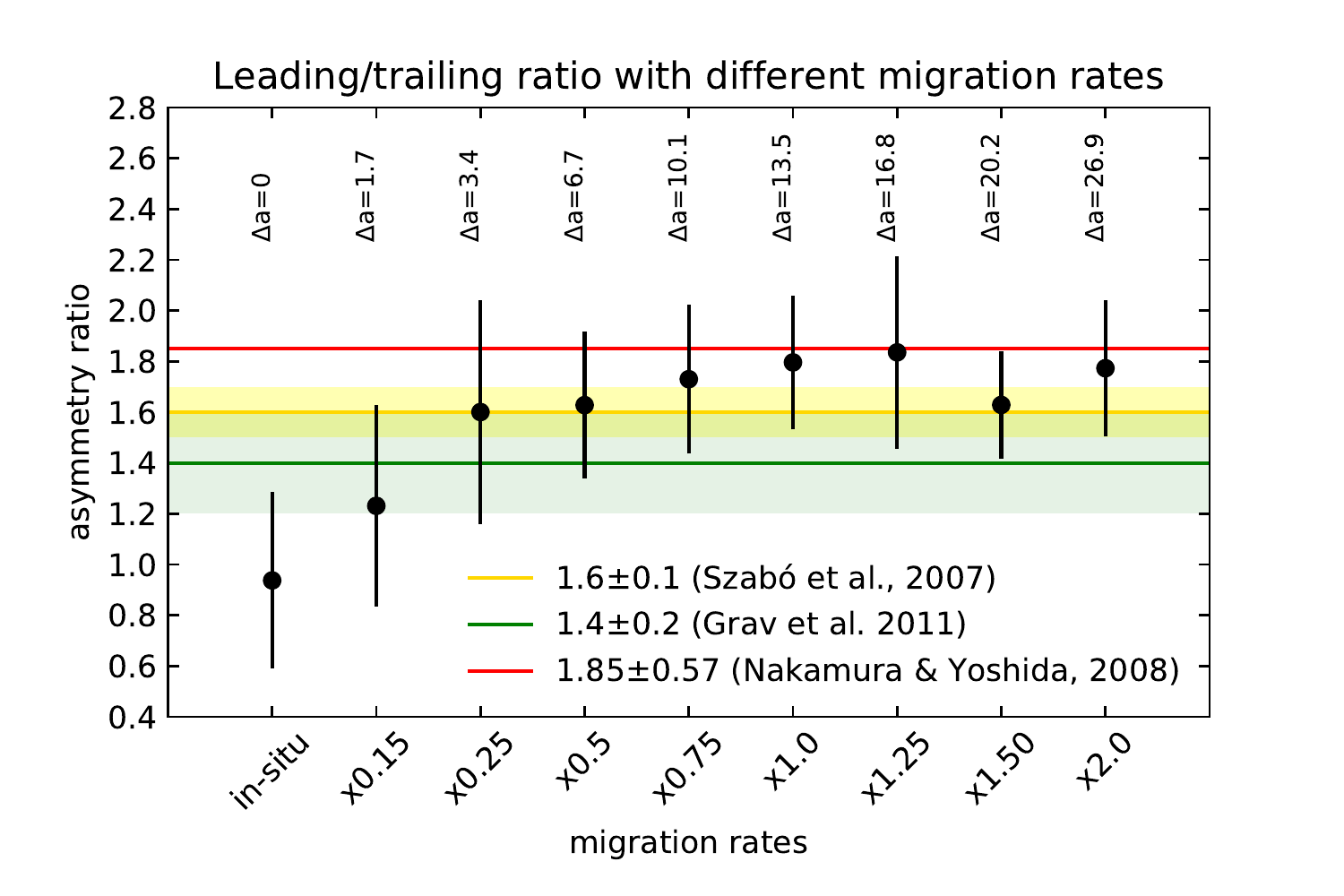}
\caption[]{The plot shows the asymmetry ratio, at $t=5$ Myr, for different migration rates of the giant planets, plus the in situ scenario. The yellow shaded region highlights the observed asymmetry ratio of $1.6 \pm 0.1$ found by \citet{szabo07} and the green shaded region represents the observed asymmetry ratio of $1.4\pm0.2$ found by \citet{grav11}. The red line indicates the estimated asymmetry ratio of $1.85\pm0.57$ found by \citet{nakamura2008b}. In this latter case the shaded region has not been overplotted to keep the figure readable, but the uncertainty is indicated in the legend. Above each asymmetry ratio it is noted the corresponding migration $\Delta$a that Jupiter undergoes.}
\label{fig:faster}
\end{center}
\end{figure} 

In addition to the nominal and the in situ model, we decided to vary the migration rate of the giant planets to check how the number of Trojans and their asymmetry ratio change due to a faster or a slower migration. The results are shown in Table \ref{table:rates}. 
The faster the migration, the more Trojans are captured. The increase in the number of Trojans captured with increasing migration rate is likely due to the larger Hill sphere at larger starting semimajor axis. Indeed, we fit $N \propto a^{1.15}$, close to the linear scaling expected purely from growth of the Hill sphere. As shown in Figure \ref{fig:faster}, each migration rate produces a certain degree of asymmetry between the number of Trojans in the two swarms, with the L$_4$ group always more populated than the L$_5$ group. However, the ratio of the asymmetry decreases abruptly for the slower migration rate, that is for short-scale migration of about 3 au or less. The ratio of the asymmetry for the other different migration rates remain of the same order within the errors. The origin of the Trojan asymmetry will be discussed in section \ref{sec:asymmetry}.

\subsubsection{Mass of the Trojan asteroids}

With our simulations, we can also make a crude estimate of the mass captured as Trojans with each different migration rate. If we take into account the Minimum Mass Solar Nebula \citep{weidenschilling77,hayashi81}, in each annular region of 1 au, we expect a mass of roughly 1 M$_\oplus$. We started with 2000 particles every 1 au, hence each particle carries $5 \times 10^{-4}$  M$_\oplus$. At $t=5$ Myr, that is 2 Myr after the early migration has ceased, the mass captured as Trojans is of the order of $10^{-2}$ M$_\oplus$ for the slowest migration rate tested and for the in situ scenario and of the order of $10^{-1}$ M$_\oplus$ for the different non-zero migration rates. Nowadays the Jupiter Trojans mass is roughly $10^{-5}$ M$_\oplus$ \citep{vinogradova15}, hence the population has to undergo an heavy mass depletion, likely imputable to planetary interactions between Jupiter and Saturn, to a possible late instability of the giant planets and/or to the presence of massive embryos in the swarms. In section \ref{sec:nice}, we tested a possible instability of the giant planets after the disc dispersal and analysed the Trojan mass depletion that it causes.

\begin{figure}
\begin{center}
\includegraphics[width=\hsize]{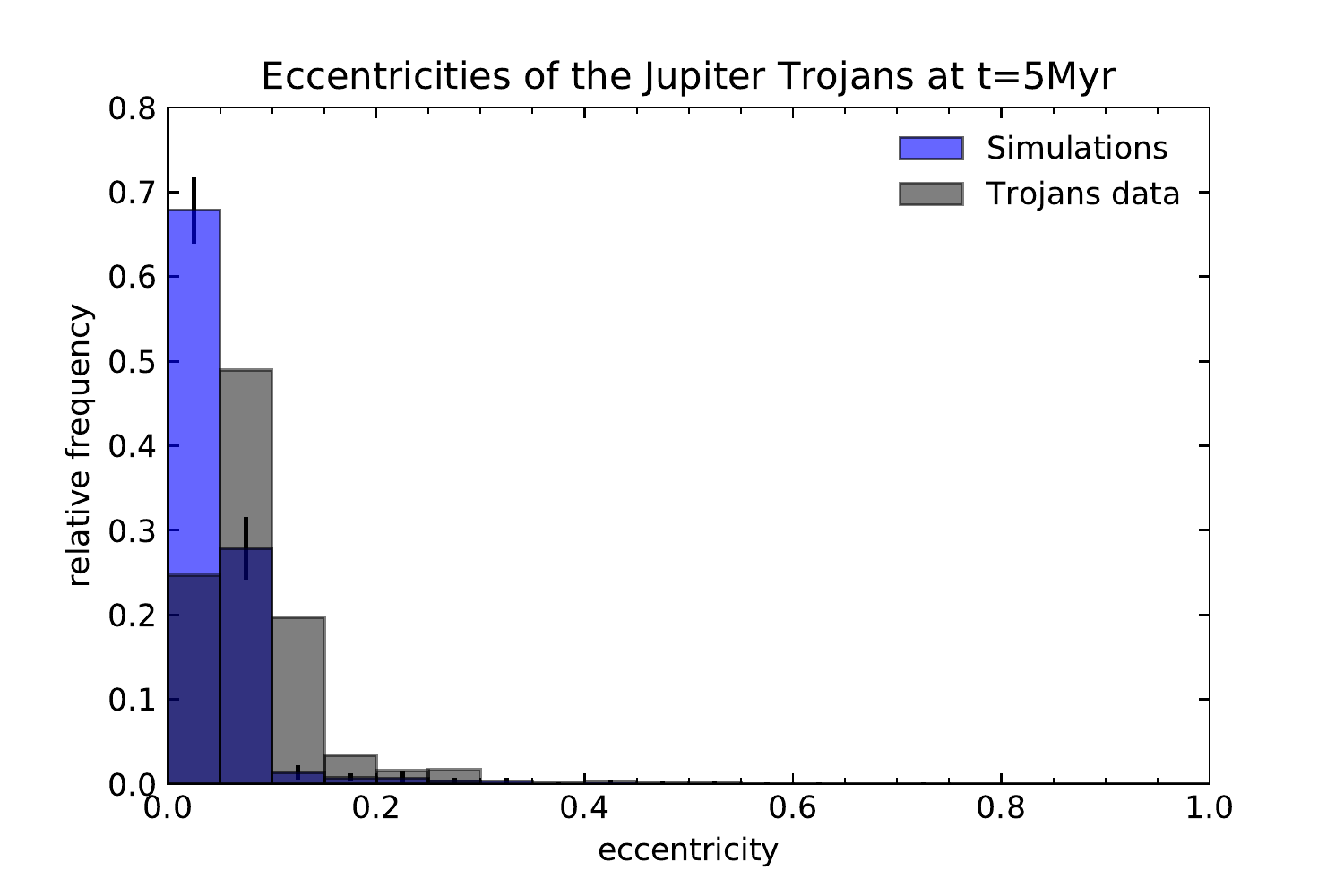}
\includegraphics[width=\hsize]{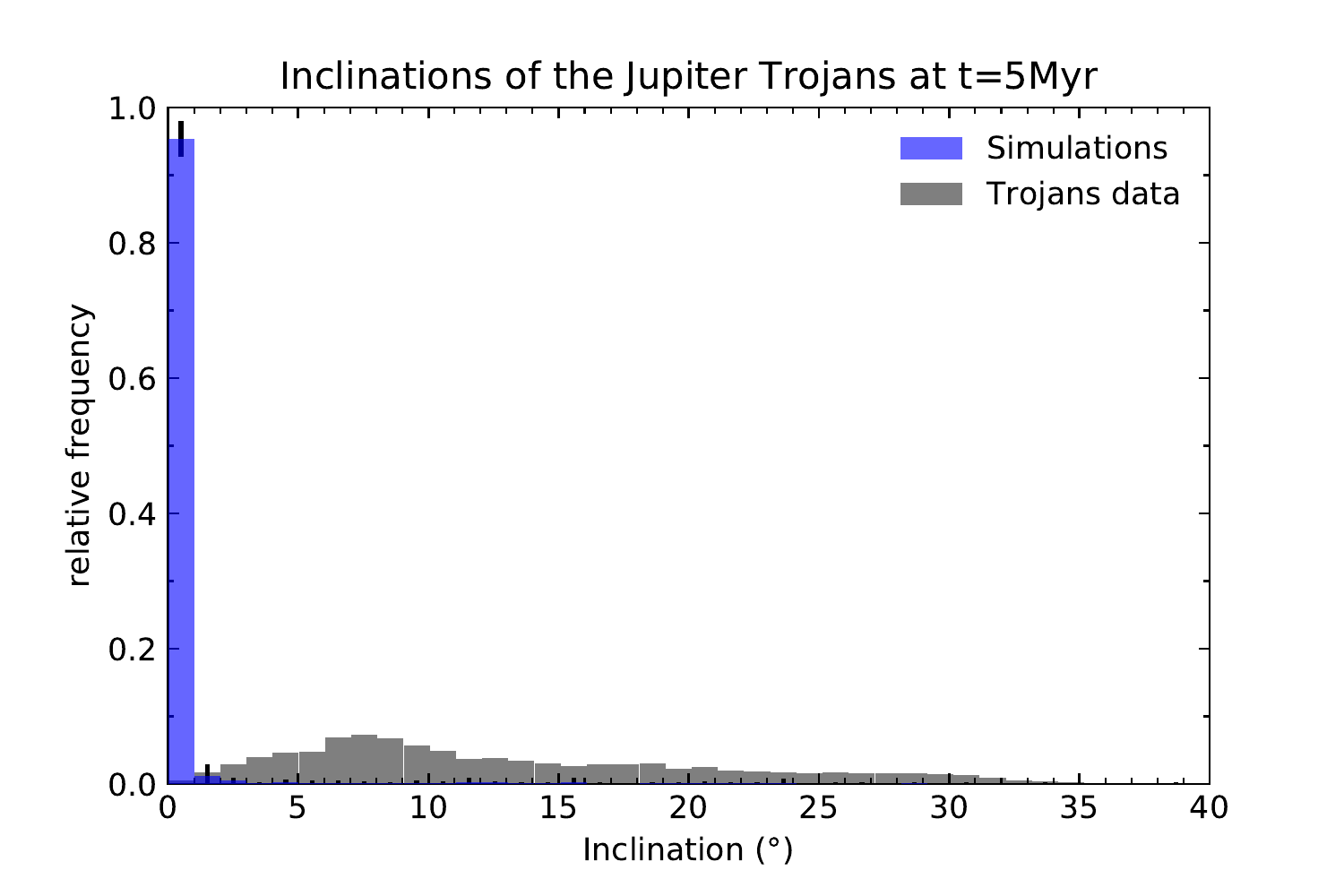}
\caption[]{Jupiter's Trojans osculating eccentricities (top figure) and inclinations (bottom figure) at $t=5$ Myr in the nominal model simulations are shown as blue histograms. The overplotted grey histograms represent the distribution of eccentricities and inclinations of the Jupiter Trojans in the  MPC database.}
\label{fig:mt}
\end{center}
\end{figure}

\subsubsection{Eccentricity and inclination distributions of the Trojans}

Figure \ref{fig:mt} shows the osculating eccentricity (top figure) and the inclination (bottom figure) distributions of the Jupiter Trojans at $t=5$ Myr, that is after 2 Myr the migration and growth has stopped. The overplotted grey histograms represent the distribution of eccentricities and inclinations of the Jupiter's Trojans in the MPC database. Comparing the results to the current distribution of eccentricities and inclinations of the Trojans, we see a moderate agreement for the eccentricities, but as regards the inclinations, we obtained a completely flat Trojan population, even if their values increased from the initial values of the order of $10^{-3}$ degrees to values of the order of $10^{-1}-10^{-2}$ degrees post-migration. This inclination issue is consistent with previous works about the in situ capture of Trojans from a growing proto-Jupiter presented in section \ref{sec:th}. 
In the in situ model, both eccentricities and inclinations of the captured Trojans remain of the order of the initial $10^{-3}-10^{-4}$ degrees.
When we tested different migration rates, the resulting eccentricity and inclination distributions are very similar to the nominal case (Figure \ref{fig:mt}).

\begin{figure}
\begin{center}
\includegraphics[width=\hsize]{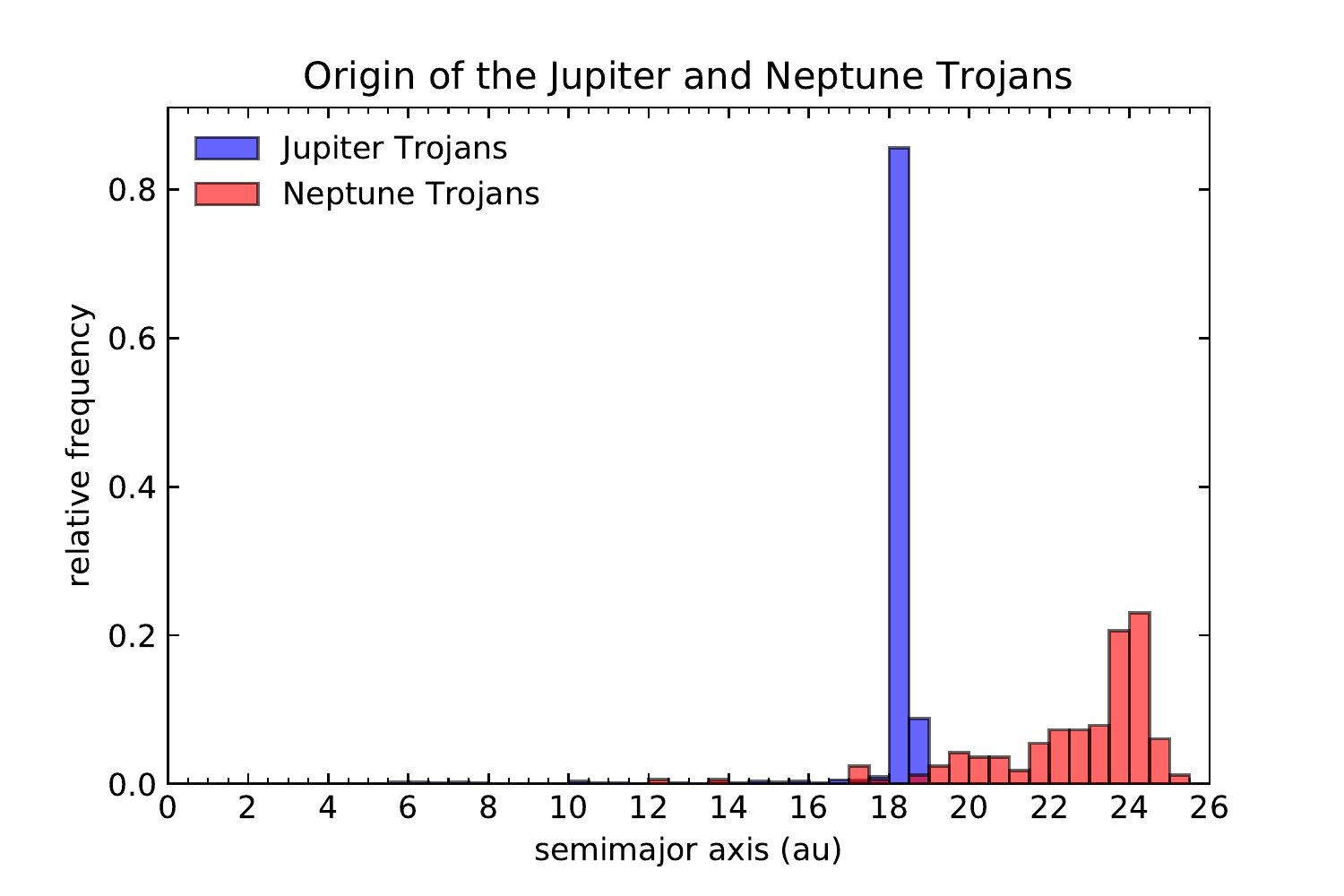}
\caption[]{Jupiter Trojans (blue histogram) and Neptune trojans (red histogram) origin in the nominal model simulations. Almost all the Jupiter Trojans are captured in the formation region of Jupiter's core at about $a\sim18$ au. Neptune Trojans are mainly particles from Neptune's core feeding zone, but also particles scattered because of the earlier growth of Jupiter and Saturn.}
\label{fig:mt2}
\end{center}
\end{figure} 

\subsubsection{Capture region of the Trojans}
The origin of the asteroids that ended up captured as Jupiter Trojans is shown in Figure \ref{fig:mt2}. Almost all the Trojan asteroids were captured in the feeding zone of Jupiter's core, at about $a\sim18$ au. Only few of them are from regions closer to the star. This result differ from the current hypothesis on the origin of the Jupiter Trojans, where they are thought to be captured in situ (close to $a\sim5$ au) or to be Kuiper Belt Objects (KBOs) implanted into the Trojans regions during the late instability of the giant planets. Our new scenario could explain the new results by \citet{jewitt18}:  the colour distribution of Neptune Trojans is statistically indistinguishable from that of the Jovian Trojans but different from the KBO's one. Indeed, in our nominal model, Jupiter's core feeding zone ($\sim18$ au) is very close to Neptune's core feeding zone ($\sim24$ au) and both of them are in a region different from the source region of KBOs. 
In the in situ model all the Jupiter Trojans condensed very close to their current location, that is between 4.5 and 5 au.
With different migration rates, the formation region of Jupiter's core varies as shown in Figure \ref{fig:rates} (and listed in Table \ref{table:difrates}) and so does the original formation region of the Jupiter Trojan asteroids.


\subsection{Origin of asymmetry}\label{sec:asymmetry}

The reason why Jupiter Trojans ended up with the L$_4$ region more populated than the L$_5$ region is caused by the relative drift between the planet and the Trojans that acts like a drag force on the asteroids. The effects of a drag force on particles in an horseshoe orbits of a planet or a satellite have been studied in many papers \citep{dermott84,murray94,murray99,sicardy03,ogilvie06}. Inward migration modifies the horseshoe orbits, increasing the width of the path of the particles in the L$_4$ side of the horseshoe orbit and decreasing the width of the particles path in the L$_5$ side of the horseshoe orbit \citep{sicardy03}. Because of this, particles spend more time on the L$_4$ side of the horseshoe orbit during the migration. This excess of L$_4$ particles becomes important when the mass growth of the planet becomes efficient in shrinking the horseshoe orbits into stable tadpole orbits \citep{fleming00}.  
\begin{figure}
\begin{center}
\includegraphics[width=\hsize]{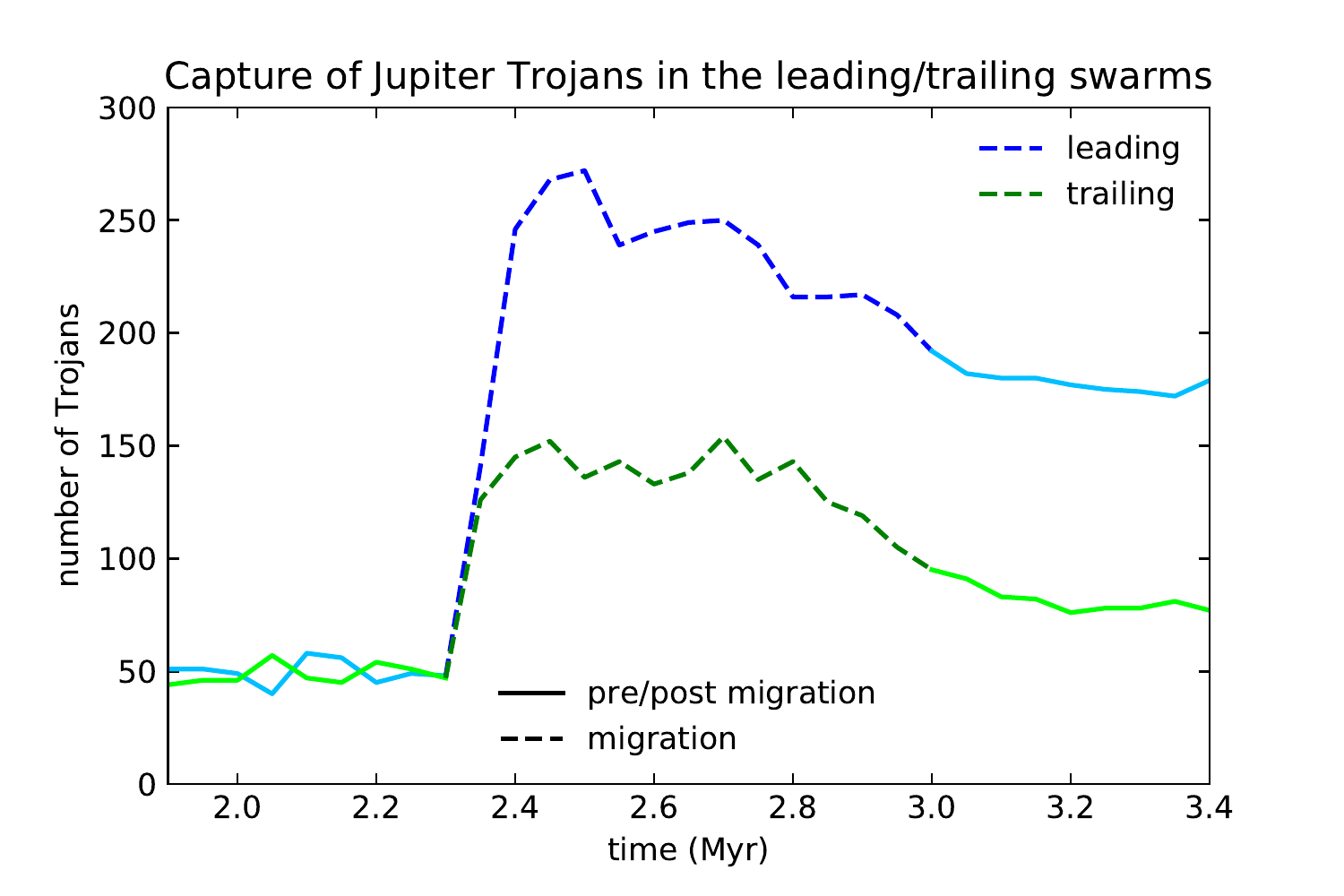}
\caption[]{The plot shows the number of particles in the leading and trailing side of the horseshoe orbits during the nominal model migration and growth of Jupiter. The blue line is the number of particles in the L$_4$ side of the horseshoe orbit (leading side) versus time and the green line is the number of particles in the L$_5$ side of the horseshoe orbit (trailing side) versus time. Horseshoe orbit encompass both L$_4$ and L$_4$, hence an excess of leading particles means that particles spend more time in the leading part of the orbit. Dashed lines indicate the time frame where Jupiter is growing and migrating.}
\label{fig:jup_troj}
\end{center}
\end{figure}
Analysing Figure \ref{fig:jup_troj}, before the migration starts ($\Delta t=0-2.3$ Myr), the horseshoe orbits are symmetric. Indeed, the number of particles in each side of the horseshoe orbit fluctuate around 50, leading to a symmetric ratio equal to 1, as also showed in Figure \ref{fig:asymmetry}, bottom plot. During the migration and growth (dashed lines), there is an excess of particles orbiting in the L$_4$ side of the horseshoe orbit because of the path of the particles in the leading side is wider than that in the trailing side and particles spend more time there. Contemporarily, the growth of the planet shrinks the horseshoe orbits into stable tadpole orbits and the asymmetry ratio is preserved. This is exactly what we found in our simulations for the Jupiter Trojans: when the migration stops, Jupiter has grown enough to be left with just tadpole orbits that preserved the asymmetry.

If the migration rate is too high L$_5$ can merge with L$_3$ or even disappear, but as shown in Figure \ref{fig:jup_troj}, L$_5$ is never completely unstable and preserved a certain amount of Trojans even when the rate of migration is doubled. So the fact that Jupiter fails to trap Trojans while migrating can not be attributed to the deformation and disappearance of one of the Trojan regions but, as we already anticipated in subsection \ref{sec:ta}, it is more likely that the sweeping of the resonances ahead of Jupiter excited the eccentricity of the bodies above the Hill eccentricity leading to an inefficient capture.

\begin{figure}
\begin{center}
\includegraphics[width=\hsize]{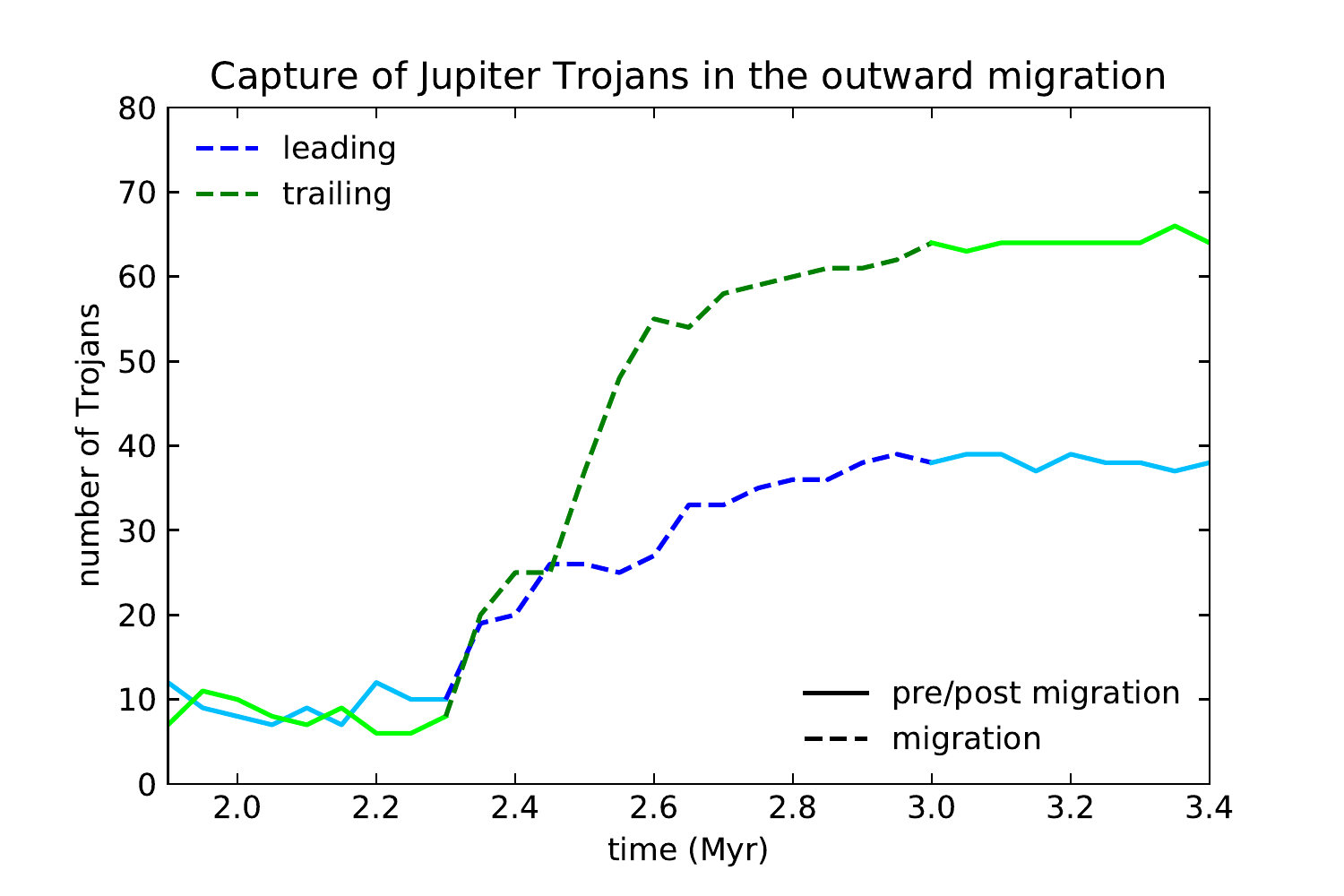}
\caption[]{The plot shows the number of particles in the leading and trailing side of the horseshoe orbits during the outward migration and growth of Jupiter from 5 au to 18 au. The blue line is the number of particles in the L$_4$ side of the horseshoe orbit (leading side) versus time and the green line is the number of particles in the L$_5$ side of the horseshoe orbit (trailing side) versus time. As expected, in the case of outward migration, particles spend more time in the trailing part of the orbit, because the path of the particles is wider on that side. Dashed lines indicate the time frame where Jupiter is growing and migrating.}
\label{fig:outward}
\end{center}
\end{figure}

In order to confirm the mechanism behind the asymmetry ratio of the Trojans, we repeated the simulations without applying any gas drag to the resonant particles and we obtained the same asymmetry, suggesting that the relative drift between the planet and the resonant particle is indeed the cause of asymmetry.

We made a test to further confirm the mechanism just described: we set a simulation with Jupiter's seed at 5 au that grows and migrates until 18 au, that is we tested an outward migration of Jupiter during its growth. In this case the deformation of the stable regions should be the opposite: in the L$_4$ side of the horseshoe orbit, particles width path will be smaller and in the L$_5$ side it will be wider, with particles spending more time in the trailing side of the horseshoe orbit. Then when the mass growth will shrink the horseshoe orbits in stable tadpole orbits we should find an opposite asymmetry, i. e. $\rm{N_{L_{4}}/N_{L_{5}}}<1$. The results of the outward migration confirm this scenario: as shown in Figure \ref{fig:outward}, we found an asymmetry ratio of $\rm{N_{L_{4}}/N_{L_{5}}}\sim0.6$ that is opposite to the one found in the inward migration.

\begin{figure}
\begin{center}
\includegraphics[width=\hsize]{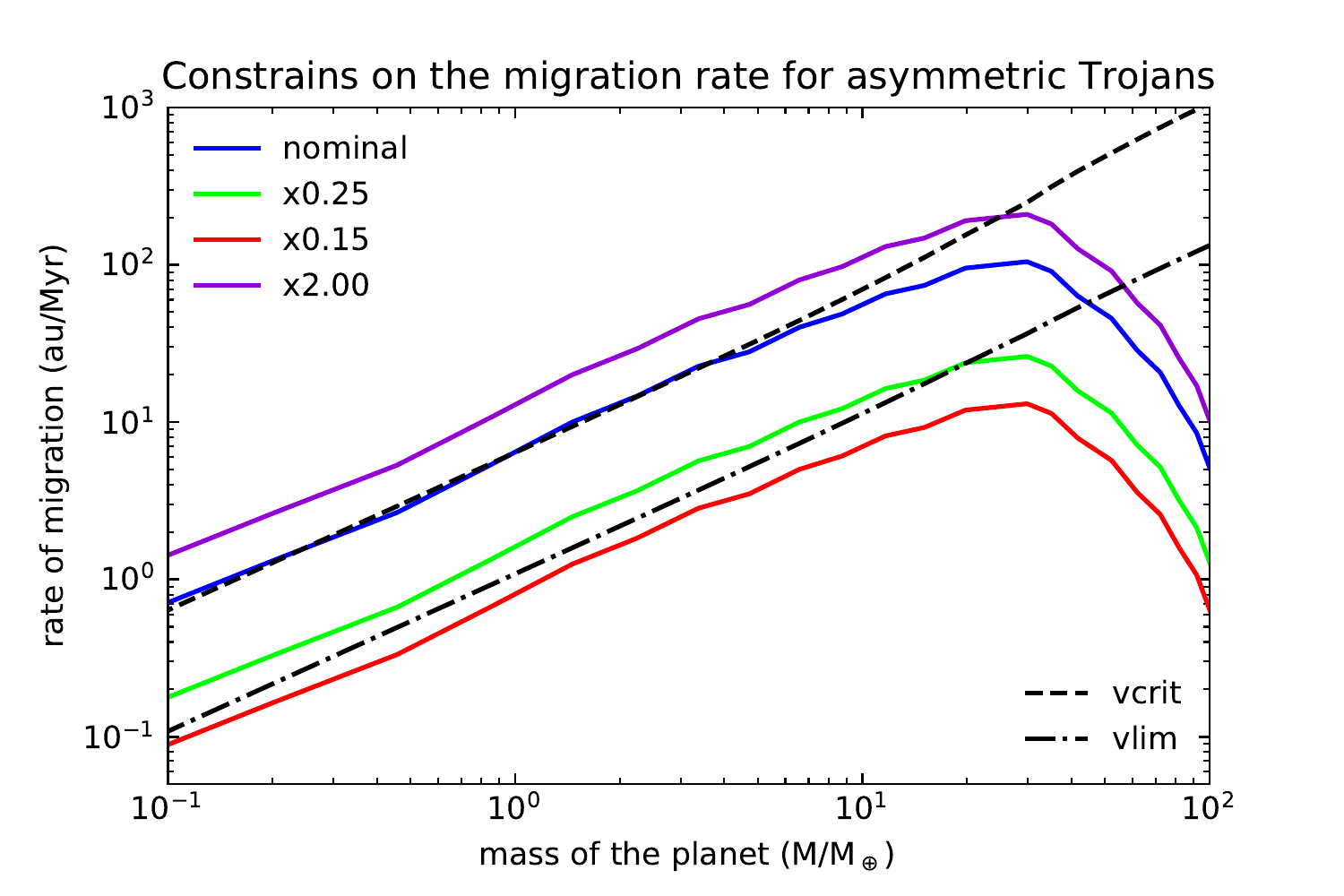}
\caption[]{Migration rates for the nominal model (blue line), the growth track 2 times faster than the nominal one (violet line) and the growth tracks 0.25 and 0.15 times the nominal one (green and red line, respectively). The black dashed line represents the critical migration rate at which L$_3$ and L$_4$ start to merge and disappear $\varv_{crit} \approx 1.45 q \Omega r$ \citep{ogilvie06}. The black dash-dotted line represents the lower limit for the migration rate to get an asymmetry comparable with the one observed and estimated in our simulations, that is $\varv_{lim}\approx0.17q\Omega r$.}
\label{fig:limit}
\end{center}
\end{figure}

The fact that for slower migration rates we found that the asymmetry ratio abruptly decreases can be a consequence of a low relative drift between the planet and the Trojans, that is a weak drag force applied to the particles. We can make an attempt to constrain the migration rate necessary to get an asymmetry ratio comparable to the one observed and estimated. With the current growth rate, when the core is 1 M$_\oplus$, a migration rate $\gtrsim 1$ au/Myr is needed to produce asymmetry (Figure \ref{fig:limit}), that corresponds to the migration rate of the growth track 0.25 times the nominal model, when the asymmetry ratio starts to decrease. An upper limit for the migration rate is provided by \citet{ogilvie06} where $\varv_{crit} \approx 1.45 q \Omega r$ represents the migration rate at which L3 and L4 start to merge and disappear (black dashed line in Figure \ref{fig:limit}). $q$ is the mass ratio between the planet and the Sun, $\Omega$ is the Keplerian angular velocity, r is the radial distance from the Sun. With our simulations, we can define a lower limit for the migration rate of the planet in order to get an asymmetry comparable with the one observed and estimated. This limit is in between the slower migration rates 0.25 and 0.15 times the nominal one, where the asymmetry decreases abruptly and is estimated to be $\varv_{lim}\approx0.17q\Omega r$ (black dash-dotted line in Figure \ref{fig:limit})
 
\begin{figure}
\begin{center}
\includegraphics[width=\hsize]{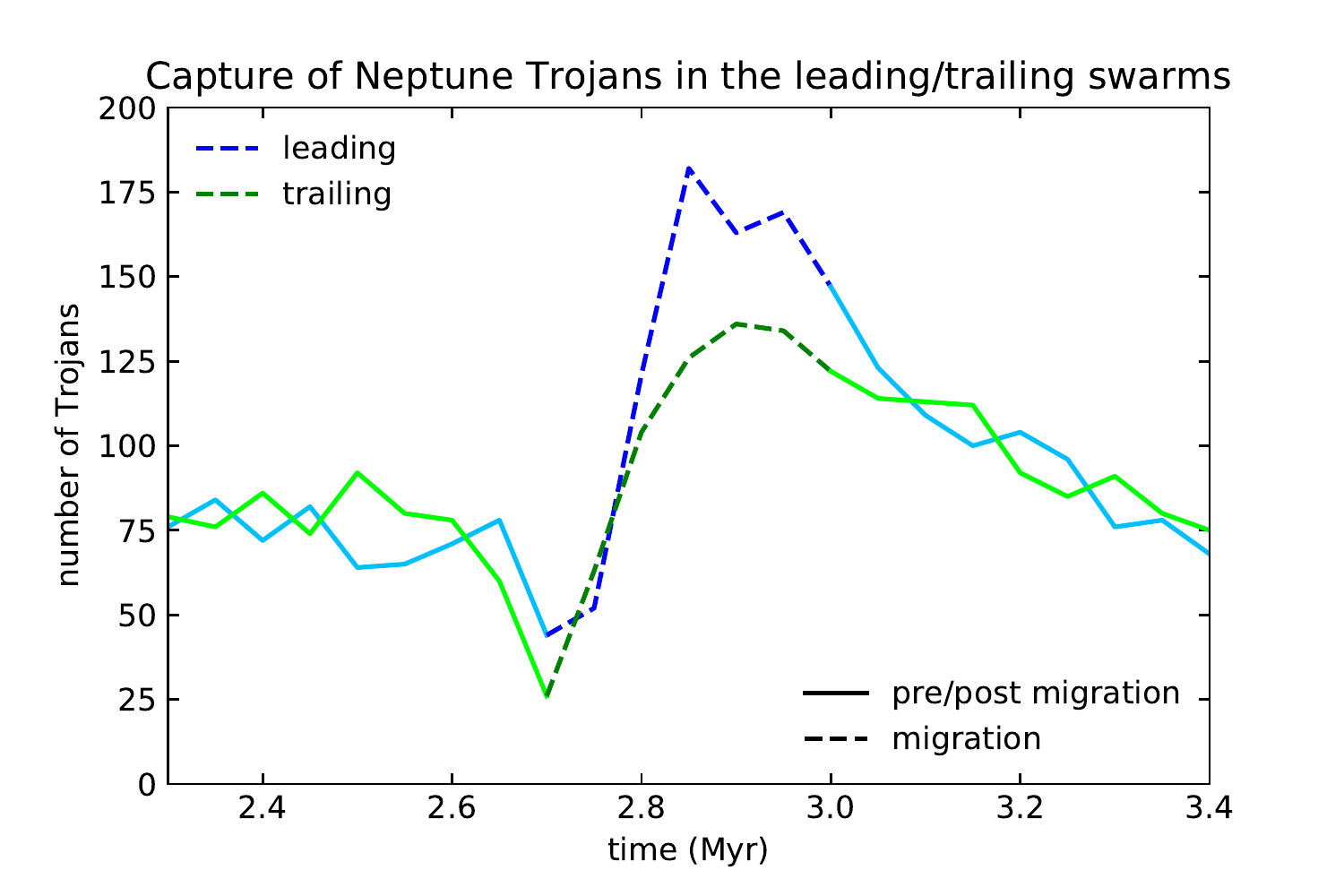}
\caption[]{The plot shows the number of particles in the leading and trailing side of the horseshoe orbits during the nominal model migration and growth of Neptune. The blue line is the number of particles in the L$_4$ side of the horseshoe orbit (leading side) versus time and the green line is the number of particles in the L$_5$ side of the horseshoe orbit (trailing side) versus time. We can clearly see an excess of leading particles during the growth and migration due to the deformations of the horseshoe orbit, but Neptune has not grown enough to shrink the orbits to stable tadpole orbits in order to preserve the asymmetry. Then when the migration stops the excess simply disappears because the horseshoe orbits are back symmetric. Dashed lines indicate the time frame where Neptune is growing and migrating.}
\label{fig:nep_troj}
\end{center}
\end{figure}

Regarding Neptune Trojans, the relative drift between the planet migrating and the Trojans produces the same excess of bodies in the L$_4$ side of the horseshoe orbits that we found for the Jupiter Trojans, as shown in Figure \ref{fig:nep_troj}. The difference, in this case, is that the mass growth of Neptune is not enough to shrink the horseshoe orbits into stable tadpole orbits as it happens for Jupiter and hence the asymmetry is not imprinted in the Trojan population. When the migration stops, particles still have horseshoe orbits and the excess in the L$_4$ side of the horseshoe orbit disappears because the path of the particles return to have a symmetric geometry. After a long term evolution, horseshoe particles are lost because of planetary interactions and Neptune is left with few original tadpole orbits that were not affected by the asymmetric capture of horseshoe particles. Our simulations clearly predict no asymmetry for the Neptune Trojans.


\subsection{Characterisation of the Hildas}

In the large-scale migration, because of the sweeping of Jupiter's resonances, asteroids are trapped in the 3:2 resonance with Jupiter (the Hilda asteroids) and in other first and second order resonances as shown in Figure \ref{fig:res}.
\begin{table}
\caption{Mean number of Hilda asteroids in the nominal model and in the in situ model simulations}             
\label{table:hilda_table}     
\centering                          
\begin{tabular}{r c c }      
\hline\hline                 
 Time (Myr) & $\rm{N_{Hildas} (migration)}$ & $\rm{N_{Hildas} (in situ)}$\\    
\hline                        
    $5$  & $1682\pm 37$ & $181\pm 11$\\
    $10$  &  $1315\pm 46$ & $174\pm 11$\\
    $50$  &  $1034\pm 51$ & $170\pm 11$\\
    $100$  & $948\pm 54$ & $166\pm 11$\\
    $200$  &  $858\pm 64$ & $156\pm 12$\\  
\hline                                  
\end{tabular}
\end{table}
In our nominal model, the evolution of the mean number of Hilda asteroids trapped in the 3:2 resonance is listed in Table \ref{table:hilda_table} and, as expected for the Hildas, their number slowly decays because the orbits are chaotic.
\begin{table}
\caption{Mean quantities of the Hilda asteroids at $t=5$ Myr, with different migration rate of the giant planets }             
\label{table:rates_h}      
\centering                          
\begin{tabular}{l c }       
\hline\hline                 
 Migration rate & $\rm{N_{tot}}$   \\    
\hline                      
    no migration  & $181\pm 11$  \\
    $\times$0.15& $1465\pm 13$\\
    $\times$0.25 & $3822\pm 13$\\
    $\times$0.50  & $2551\pm 40$ \\
    $\times$0.75  & $1762\pm 27$  \\
    $\times$1.00& $1682\pm 37$ \\
    $\times$1.25  & $1843\pm 41$  \\      
    $\times$1.50  & $1140\pm 40$  \\
    $\times$2.00 & $375\pm 16$  \\ 
\hline                                  
\end{tabular}
\end{table}
Table \ref{table:rates_h} shows the number of Hilda asteroids trapped in the in situ growth of Jupiter and with different migration rates at $t=5$ Myr, that is after 2 years that the growth and the migration of the giant planets has ceased. The capture efficiency has a maximum for the slower migration rates and decreases for faster migration rates.

\subsubsection{Mass of the Hildas} 

To estimate the mass that ends up in the 3:2 resonance with Jupiter, we can use the same method used to compute the mass of the Jupiter Trojans. Taking into account the Minimum Mass Solar Nebula, each particle carries $5 \times 10^{-4}$  M$_\oplus$. The mass captured in the 3:2 resonance is of the order of $1$ M$_\oplus$ for the nominal model, with a maximum for the migration rate 0.25 times slower that trapped almost $2$ M$_\oplus$ in the resonance and a minimum for the fastest migration rate ($\times 2$) with roughly $0.2$ M$_\oplus$ captured in the resonance. In the in situ growth model, roughly $0.1$ M$_\oplus$ ended up in the 3:2 resonance. As for the Jupiter Trojans case, the Hilda group needs to undergo a heavy mass depletion in order to match the current mass likely because of unstable chaotic orbits in the resonance and because of a possible late instability of the giant planets as discussed in section \ref{sec:nice}.

\subsubsection{Eccentricity and inclination distributions of the Hildas} 
\begin{figure}
\begin{center}
\includegraphics[width=\hsize]{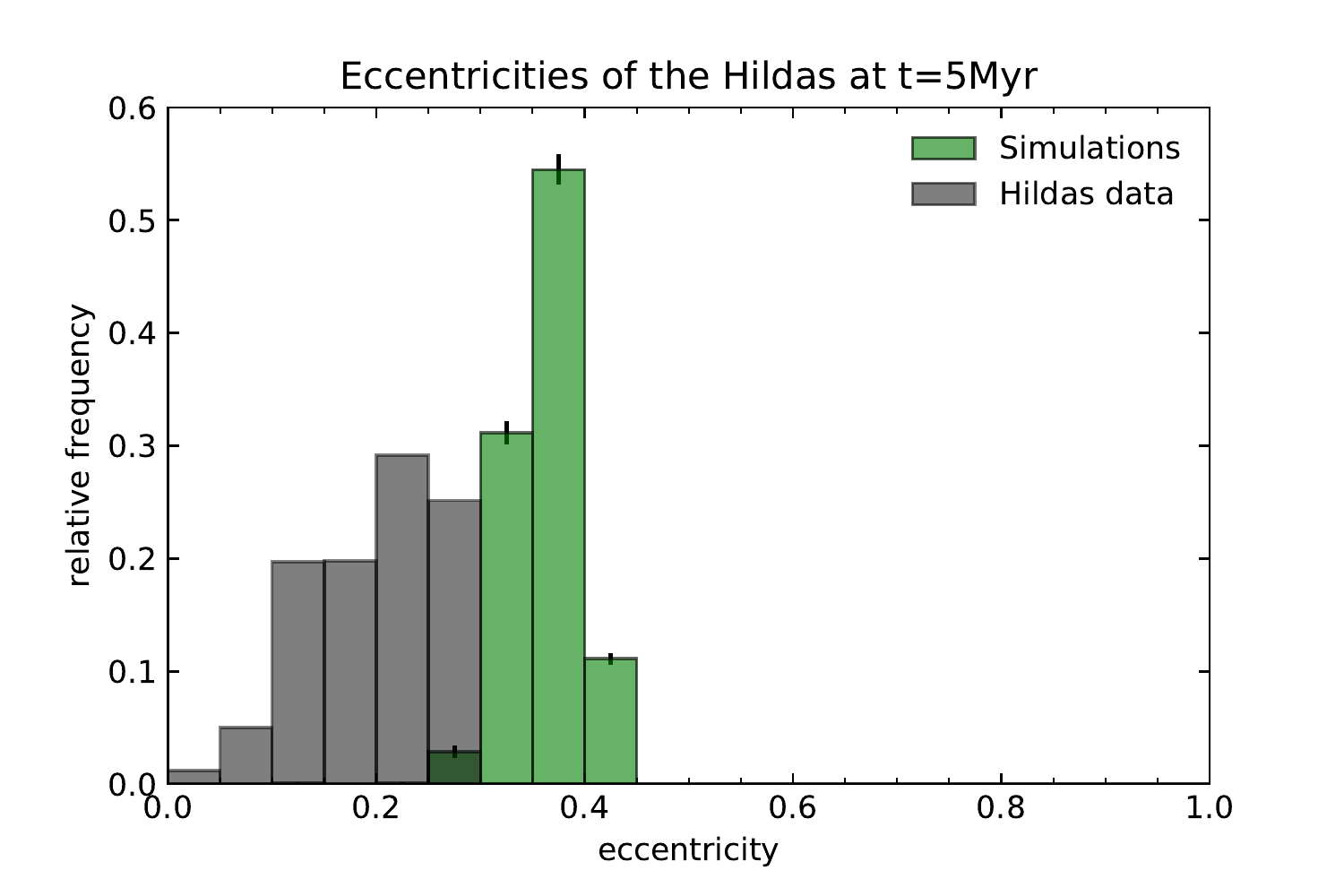}
\includegraphics[width=\hsize]{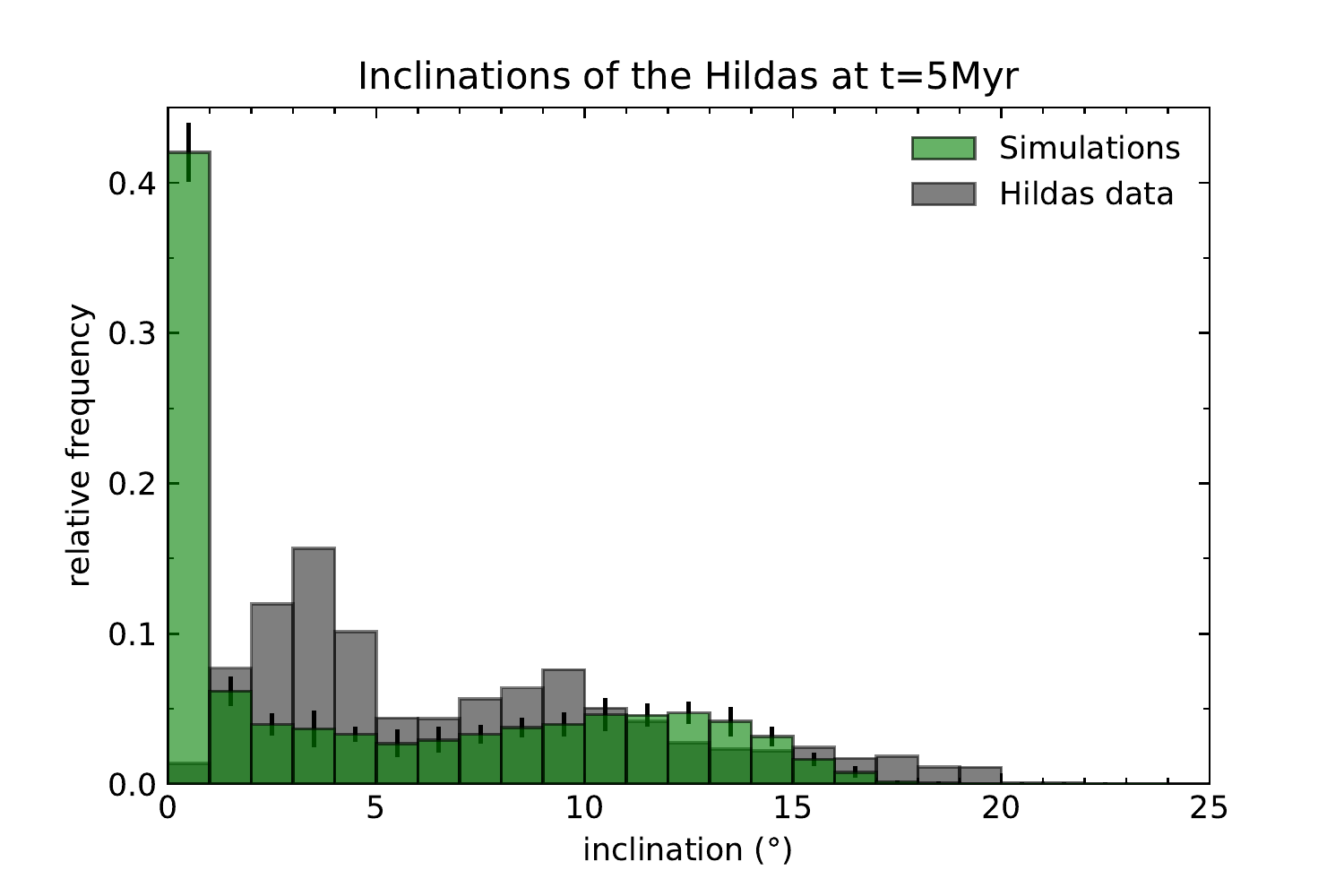}
\caption[]{Osculating eccentricities (top figure) and inclinations (bottom figure) of the Hilda asteroids at $t=5$ Myr with Jupiter and Saturn that end locked in the 1:2 resonance.}
\label{fig:mh0}
\end{center}
\end{figure}
Figure \ref{fig:mh0} shows the osculating eccentricities of the Hilda asteroids (top figure) after the nominal model migration with Jupiter and Saturn ending in their 1:2 resonance. 
The Hilda asteroids are characterised by a high eccentricity with a peak between $e=0.3$ and $0.4$, with a lack of low-eccentricity bodies and by inclinations up to $18^\circ$ as shown in the bottom histogram of Figure \ref{fig:mh0}. Implications of the high eccentricities of the Hildas in this phase will be discussed in section \ref{sec:nice}. As we can see from the time snapshots in Figure \ref{fig:ploti100}, the inclinations of the Hildas are excited by the migration of Saturn relative to Jupiter, hence we expect that in the in situ scenario inclinations are not excited. 
\begin{figure}
\begin{center}
\includegraphics[width=\hsize]{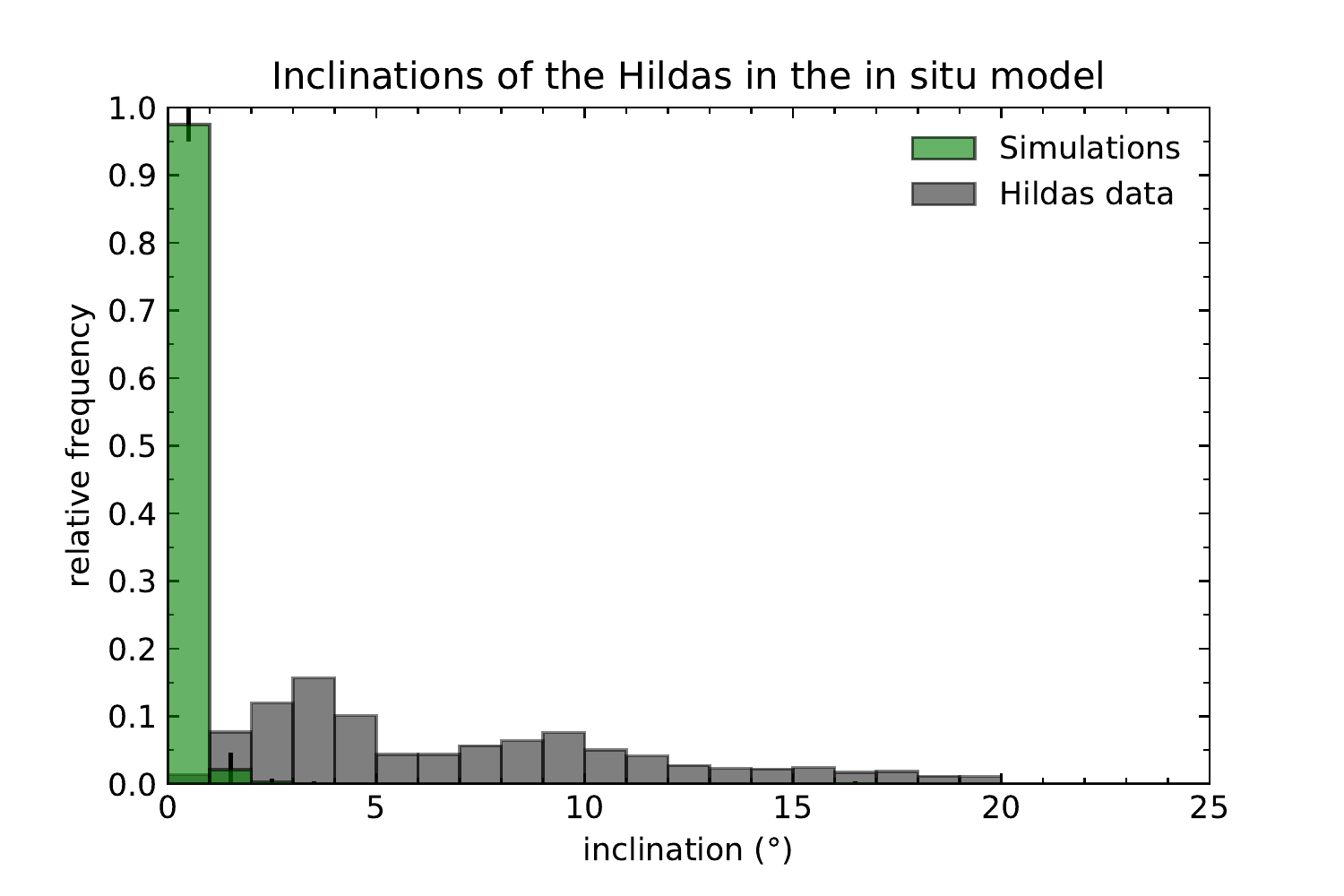}
\caption[]{The green histogram indicates the osculating inclination distribution of the Hilda asteroids in the in situ growth model. The grey histogram indicates the osculating inclination distribution of the Hildas in the MPC database.}
\label{fig:hildai}
\end{center}
\end{figure}
Results are shown in Figure \ref{fig:hildai}: in the in situ model, Hildas have lower eccentricities (less then 0.1) and a very flat inclination distribution. We also found a completely flat distribution of the inclinations for slow migration rates (0.50, 0.25 and 0.15 times slower than the nominal model) because Saturn starts too close to Jupiter and for faster migration rates (1.50, 2.00 faster than the nominal model) because Saturn migrates too fast and the passage of Saturn's resonances have no time to excite the Hilda asteroids.
The subsequent evolution of the inclinations and eccentricities of the Hilda asteroids is discussed in section \ref{sec:nice}.

\subsubsection{Capture region of the Hildas}
\begin{figure}
\begin{center}
\includegraphics[width=\hsize]{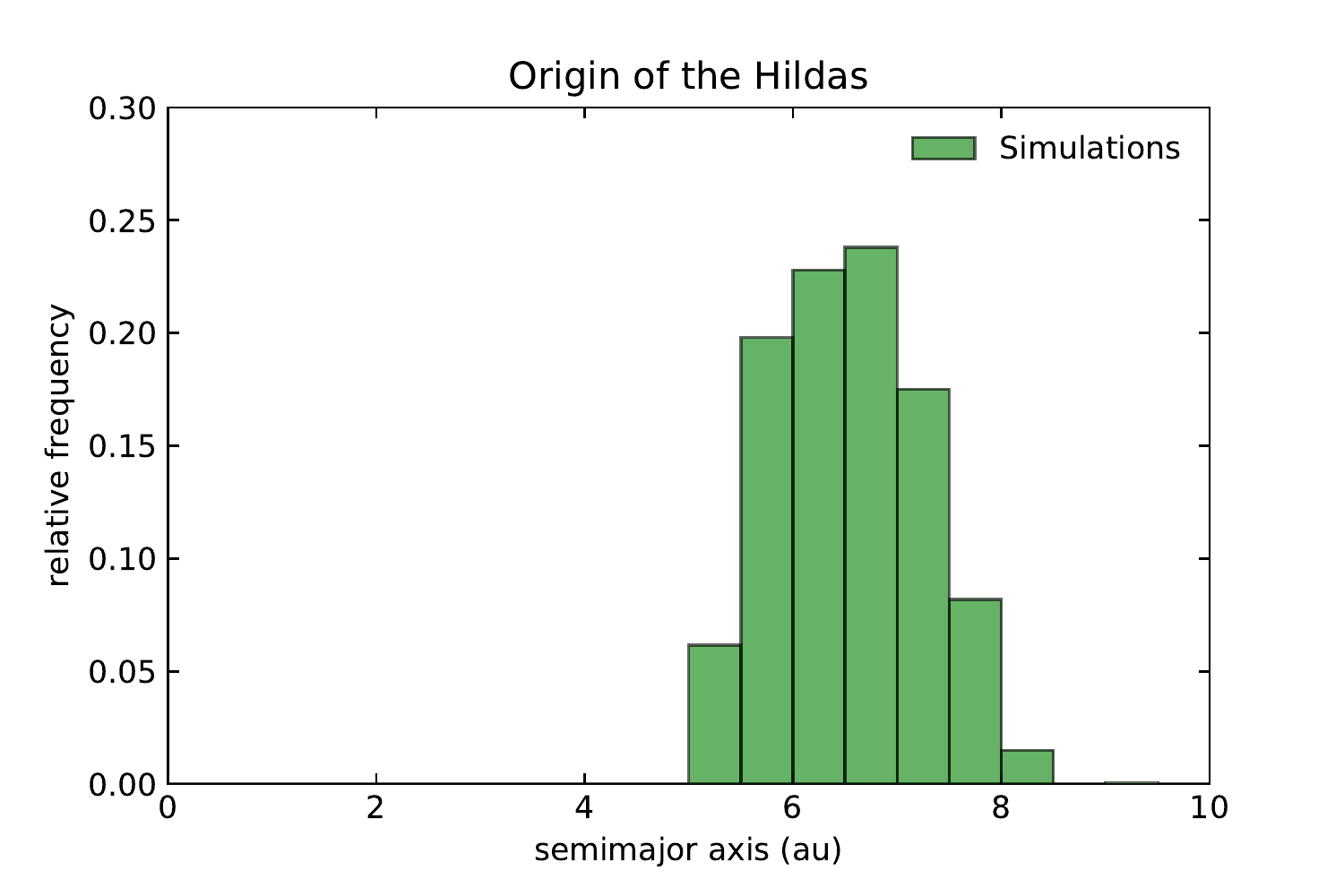}
\caption[]{Original formation regions of the bodies trapped in the 3:2 mean-motion resonance with Jupiter.}
\label{fig:mh}
\end{center}
\end{figure}
Figure \ref{fig:mh} shows the origin of the objects trapped in the 3:2 resonance in our nominal model: the Hilda asteroids are roughly formed between 5 to 8 au, with a peak around 7 au. 
\begin{figure}
\begin{center}
\includegraphics[width=\hsize]{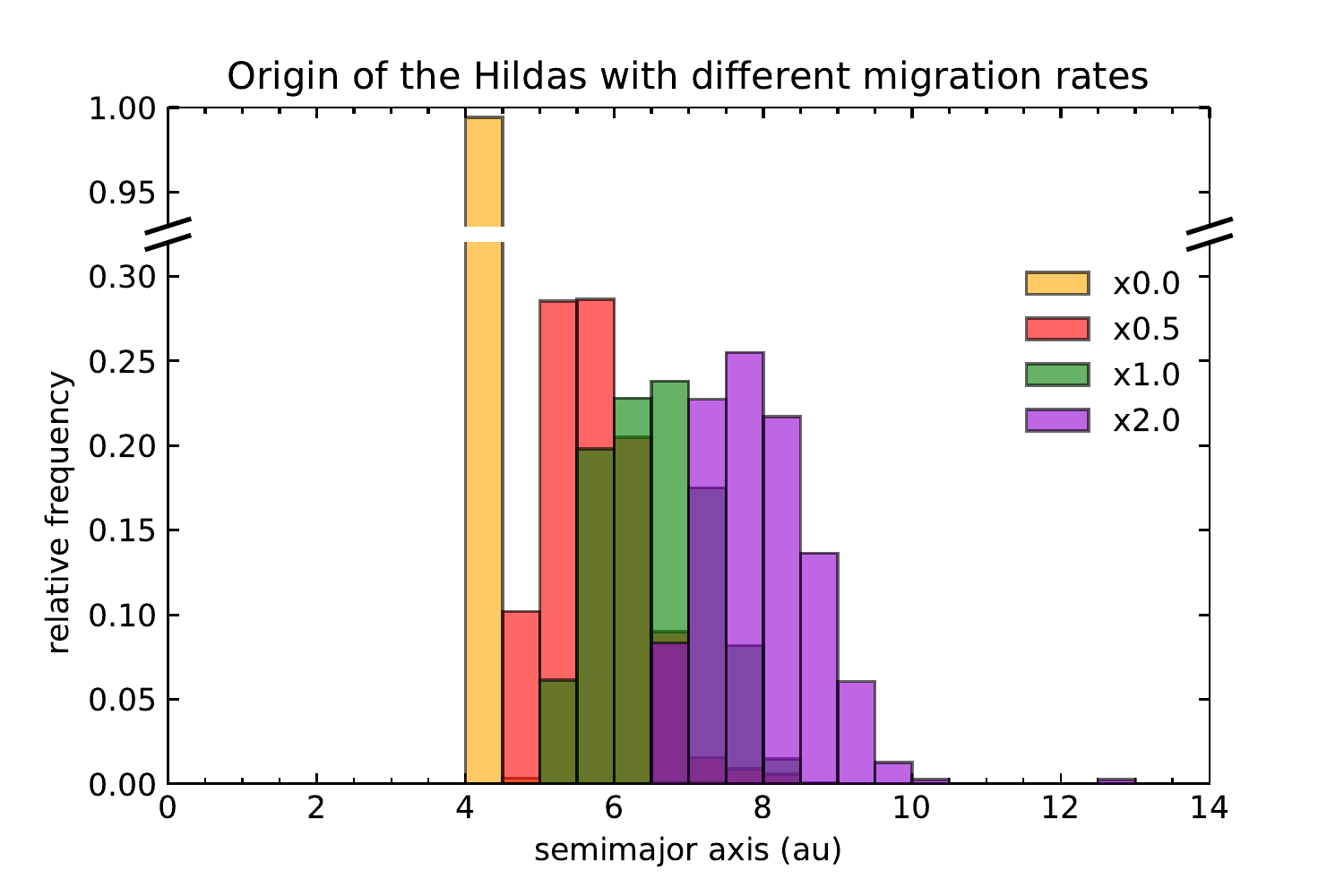}
\caption[]{Origin of the Hilda asteroids with different migration rates. The yellow histogram is referred to the in situ growth model, the red histogram corresponds to a migration rate 0.5 times slower that the nominal one. The green one is the result for the nominal model and the purple one corresponds to a migration rate 2 times faster than the nominal model.}
\label{fig:mh2}
\end{center}
\end{figure}
However, depending on the specific rate of migration, Hilda asteroids are captured in slightly different locations as shown in Figure \ref{fig:mh2}. With the faster migration rates, the histogram is just displaced towards larger semimajor axes by about 1.5 au and the peak moves from 7 to 8 au. If the migration rate is slower, the peak moves from 7 to 5 au, with the extreme case of the in situ growth where the bodies all come from the 4 to 4.5 au annular region.


\subsection{The asteroid belt}

In our nominal model simulations, we found that particles that originally formed from 4 to 16 au are implanted in the inner solar system. Part of them ended up in the asteroid belt and part of them ended up in the terrestrial planet region or posses eccentricity values high enough to cross the terrestrial planets region anyway. These results have important implications for the delivery of water to the asteroid belt and on Earth, as also suggested by previous studies focusing on the in situ formation or a more limited migration of the giant planets \citep{turrini11,turrini14b,grazier14}.
\citet{raymond17} simulated both an in situ growth and a large scale migration of the giant planets including gas drag. They found that a fraction of objects from beyond the current position of Jupiter are implanted in the main asteroid belt during the growth of the giant planets, generating a large-scale mixing of the small bodies. The source region of implanted asteroids was from 4 to 9 au in the in situ scenario, identified as the precursors of the C-type asteroids, and from 4 to 15 when migration was taken into account. 

In our migration model, we also found implanted bodies from 4 to 16 au in the asteroid belt, but most of them, 96.6\%, come from the 4-5 au region (that end up piled up in the 2:1 resonance with Jupiter at about 3 au as shown in Figures \ref{fig:plote100} and \ref{fig:ploti100}) and just 3.4\% of them come from 6-16 au. Differently from \citet{raymond17}, we did not simulate the preexisting asteroid belt for computational time reasons, but we can analyse our results considering a preexisting `virtual' asteroid belt composed by S- and C- type asteroids, with the S-type dominating the inner belt and C-types more common in the outer belt. Taking into account the Minimum Mass Solar Nebula, the asteroid belt mass is about 1 M$_\oplus$.
From the same assumption, a particle in our simulations carries $5\times 10^{-5}$ M$_\oplus$ and thus we can make a crude estimation of the mass that is injected in the asteroid belt. From the 5 Myr snapshots we counted the particles that ended up in the asteroid belt. The results are that in the nominal model, roughly $1$ M$_\oplus$ of outer solar system planetesimals are injected into the asteroid belt. This number is to be compared to the amount of mass in the primordial asteroid belt, that is roughly $1$ M$_\oplus$. Hence, after the migration, about 50\% of the belt is composed of outer solar system planetesimals. The nominal model is the most efficient in injecting bodies into the asteroid belt and the least efficient ones are the in situ model and the growth tracks $2.0$ times faster than the nominal model: both inject only $\sim$0.2 M$_\oplus$. 

It is important to notice that, in this model, 96.6\% of the mass injected in the belt resides in the 2:1 resonance and a subsequent instability of the giant planets after the disc dispersal can affect dramatically this resonant region, depleting it almost completely. We will show this result in section \ref{sec:nice}. This will reduce drastically the contamination of the belt, leaving just $10^{-2}$ M$_\oplus$ of P and D type asteroids in the belt to be compared with the $1$ M$_\oplus$ preexisting belt. 

Even though the contamination is not so important in terms of mass, secular erosion of the mass of the asteroid belt only accounts for 50\% of the its mass loss \citep{minton10}. In order to match its current mass, the asteroid belt has to lose another 99.9\%. This is a longstanding problem in the solar system field and many mechanisms has been proposed in order to explain the low mass of the asteroid belt, such as the presence of planetary embryos in the belt \citep{wetherill92,petit98,chambers01,petit01,bottke05b,obrien07}, the Grand Tack scenario \citep{walsh11} and the empty asteroid belt hypothesis \citep{raymond17b}. We are not going to explore further this issue since we did not simulate the inner part of the solar system and to solve it is beyond the scope of this paper.

As regards the implantation efficiency and the final distribution of implanted bodies in the asteroid belt and terrestrial planet region, they depend on the aerodynamical gas drag that damps particle eccentricities and inclinations and hence on the disc parameters \citep{raymond17}.


\section{After the disc dispersal}\label{sec:nice}

Despite the fact that the common outcome of hydrodynamical and N-body simulations of growing and migrating planets through protoplanetary discs is to end up with planets in resonances or in chains of resonance, nowadays, our giant planets are not in resonance. Studies about the resonant Trans Neptunian Objects \citep{malhotra93,malhotra95} suggested that Neptune could have migrated outwards for several au. Moreover, the shape of the gaps in the asteroid belt is consistent with the sweeping of gravitational resonances during a possible gas giants' migration \citep{minton09}. These works suggest that it is likely that a possible rearrangement of the semimajor axes of the giant planets has occurred after their early migration and growth, probably just after the disc disperses, when the damping effect of the gas is gone.

A possible instability of the giant planets has been studied by many authors \citep{fernandez84,minton09,gomes05,morbidelli05,tsiganis05,morbidelli10,levison11,nesvorny11} and it can involve different mechanisms such as planetesimal driven migration and planet-planet scattering.
 
Our aim is to investigate what effects an instability of the giant planets could have had on our Trojan and Hilda asteroids. 
In order to do so, we followed the approach adopted in \citet{pirani16} and, just after the disc dispersal ($t=3$ Myr), we let Jupiter and Saturn migrate following the exponential law provided by \citet{minton09}:
\begin{equation}\label{eq:mig}
a(t)=a_0+\Delta a[1-\rm{exp}(-t/\tau)]
\end{equation}
where $\rm{a_0}$ is the initial semimajor axis, $\Delta a$ is the final displacement. We placed Jupiter and Saturn's seeds in our nominal model in order that after the early inward migration we get a compact configuration where $a_J\approx5.4$ and $a_S\approx8.6$, respectively. The final displacement for the subsequent instability was set to  $\Delta a_J=-0.2$ for Jupiter and $\Delta a_S=0.9$ for Saturn, in order for the gas giants to reach roughly their current semimajor axis. $\tau$ is the migration e-folding time. We are interested in the survivability of the Trojans and the Hildas, so we explore different scenarios for the instability after the disc dispersal. In order to do so, we tested $\tau=0.5$ Myr \citep{minton09} to mimic a planetesimal-driven migration, though the model has some issues in reproducing the orbital properties of the main belt, as discussed in \citep{morbidelli10} and \citep{pirani16}. We also used $\tau=5$ kyr (that is a very short time-scale) to mimic a planet-planet scattering migration such as the one simulated in \citet{morbidelli10}. While planet-planet scattering is not associated with a smooth migration, due to the adopted short time scale the exponential law in Eq. \ref{eq:mig} should be a reasonable first-order approximation. As Jupiter's semimajor axis, in a planet-planet scattering, evolves in discrete steps, to be sure about the survivability of the Trojans, we also simulated the extreme case of a single sudden displacement of Jupiter's semimajor axis of 0.2 au (eccentricity and inclination also acquire instantaneously their current values). The real behaviour will be something in between the very fast smooth migration with $\tau=5$ kyr and the large instantaneous jump. 

Since we know that Trojans and Hildas come from narrow regions of the disc, we conducted separate simulations, populating just the region around Jupiter's core for the Trojans and the region between 4 and 9 au for the Hildas, in order to perform shorter integrations without losing any information.
For the eccentricities and inclinations, in the case of smooth migrations, we adopted analogous laws:
\begin{equation}
e(t)=e_0+\Delta e[1-\rm{exp}(-t/\tau)]
\end{equation}
\begin{equation}
i(t)=i_0+\Delta i[1-\rm{exp}(-t/\tau)]
\end{equation}
where $e_{\rm{0}}$ and $i_{\rm{0}}$ are the eccentricities and the inclinations at $t=3$ Myr, when the disc disperses. $\Delta e$ and $\Delta i$ are their differences from the current values.

We found that in all the migration models, the asymmetry ratio does not change significantly. The number of Jupiter Trojans keeps slowly decaying without any drastic change in the rate in case of smooth migrations, even with a very short time scale. In the extreme case of a single jump, the survivability of the Trojans depends on the displacement of Jupiter's semimajor axis. We tested $\Delta a=0.16$ au where 39.9\% of the Trojans survived and $\Delta a=0.23$ au where 20.1\% of the Trojans survived. This is an expected result because the Trojans librate in regions that are as wide as about 2 Jupiter's Hill radii (roughly 0.7 au) and, in order to lose all the Trojans, Jupiter's has to jump, in a single event, a large fraction of 1 au. This is an important result, because it means that the asymmetry ratio is primordial and has a direct link with the migration rate of Jupiter's core while it is growing. 

In all the migration model tested, we found the same results in terms of the shapes of the eccentricity and inclination distributions of the Hilda and the Trojan asteroids.
\begin{figure}
\begin{center}
\includegraphics[width=\hsize]{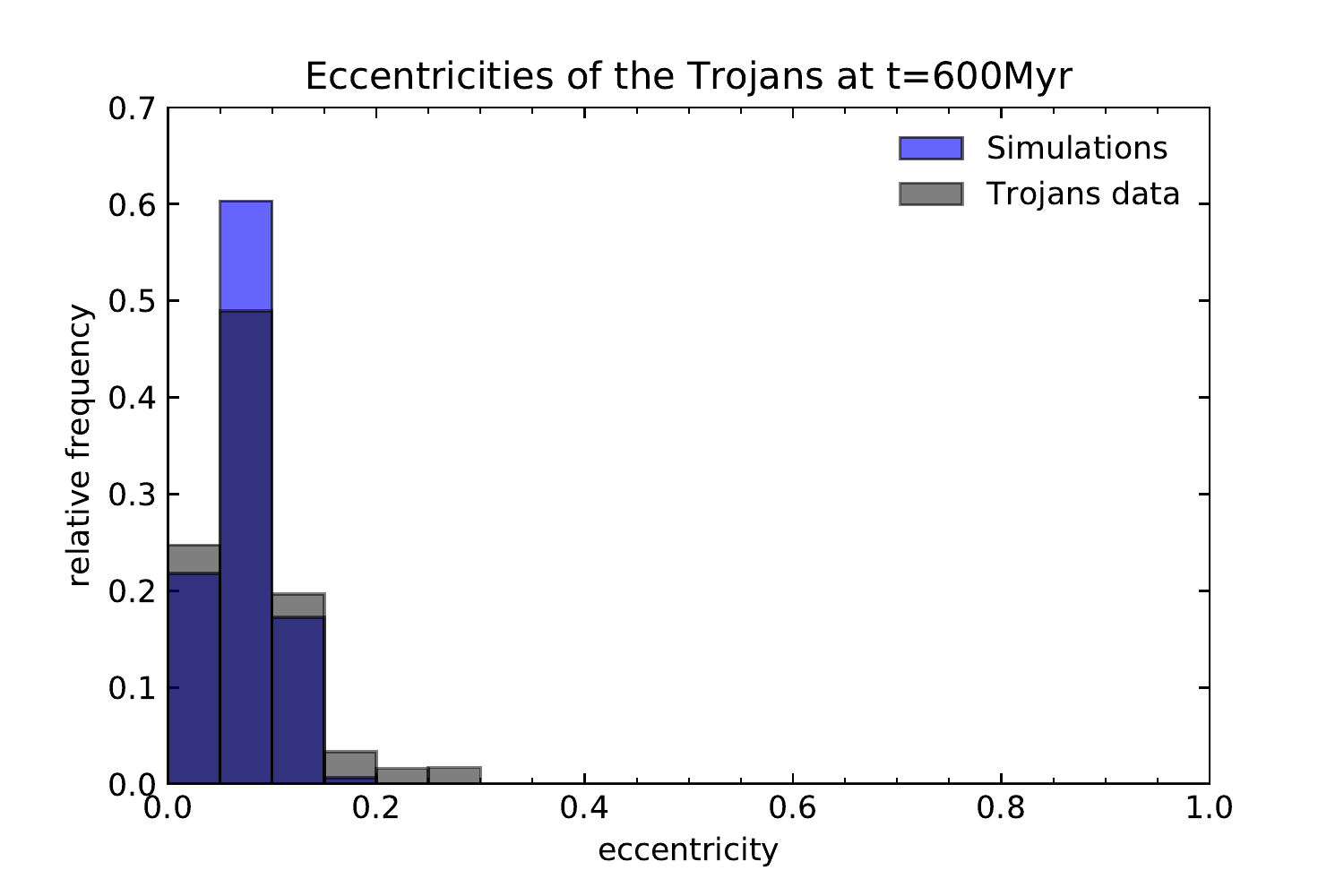}
\includegraphics[width=\hsize]{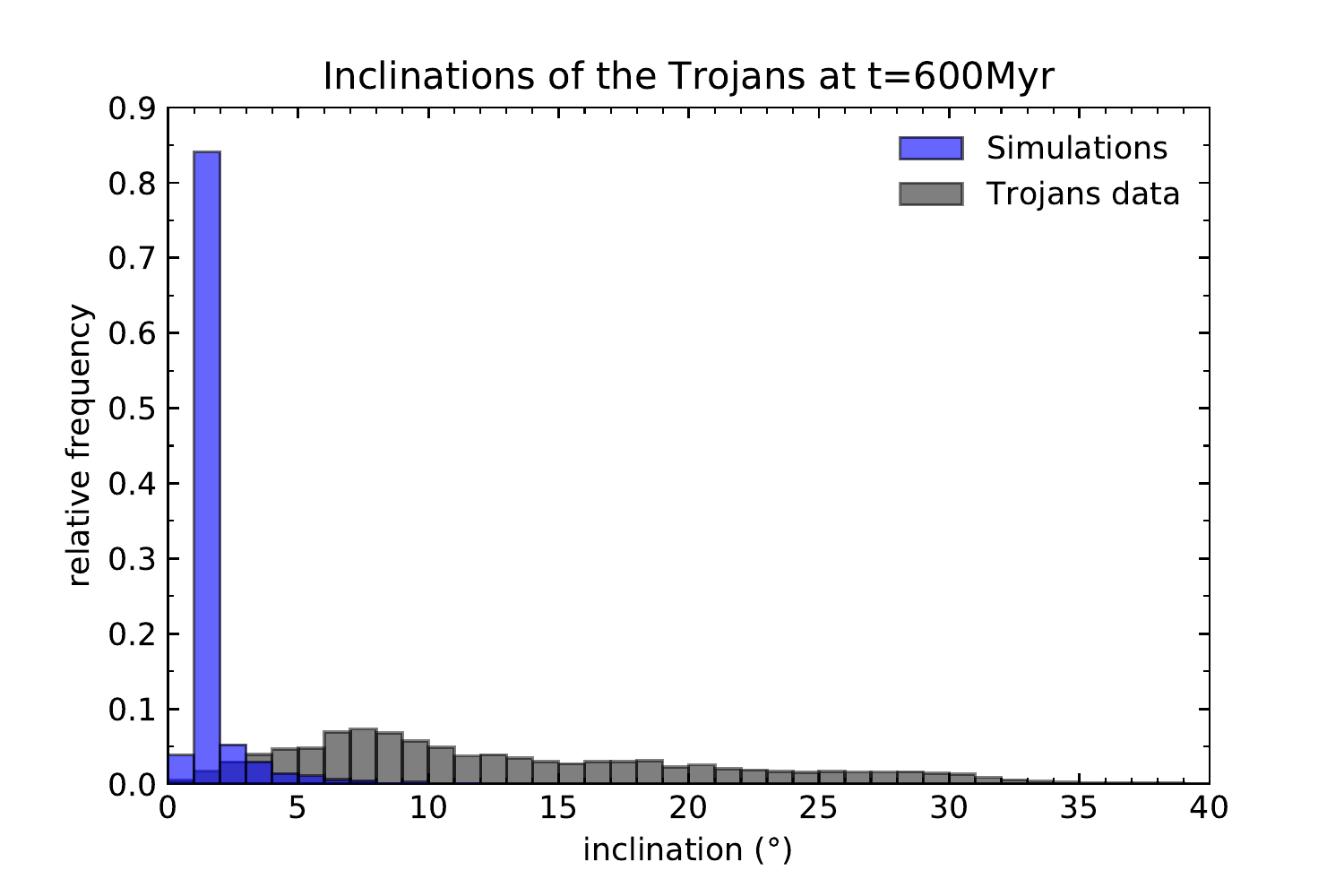}
\caption[]{Osculating eccentricities (top figure) and inclinations (bottom figure) of the Trojan asteroids at $t=600$ Myr for $\tau=0.5$ Myr when Jupiter and Saturn experience an instability after the disc disperses at $t=3$ Myr.}
\label{fig:nicetrojans}
\end{center}
\end{figure}
The resulting eccentricity and inclination distribution of the Trojans after  $t=600$ Myr for $\tau=0.5$ Myr is shown in Figure \ref{fig:nicetrojans}. While the eccentricity distribution is compatible with the current one, the inclinations of the Trojans increased from values of the order of $10^{-1}-10^{-2}$ to around 1$^\circ$, but maintained a very flat distribution compared to the current one. 

\begin{figure} \begin{center}
\includegraphics[width=\hsize]{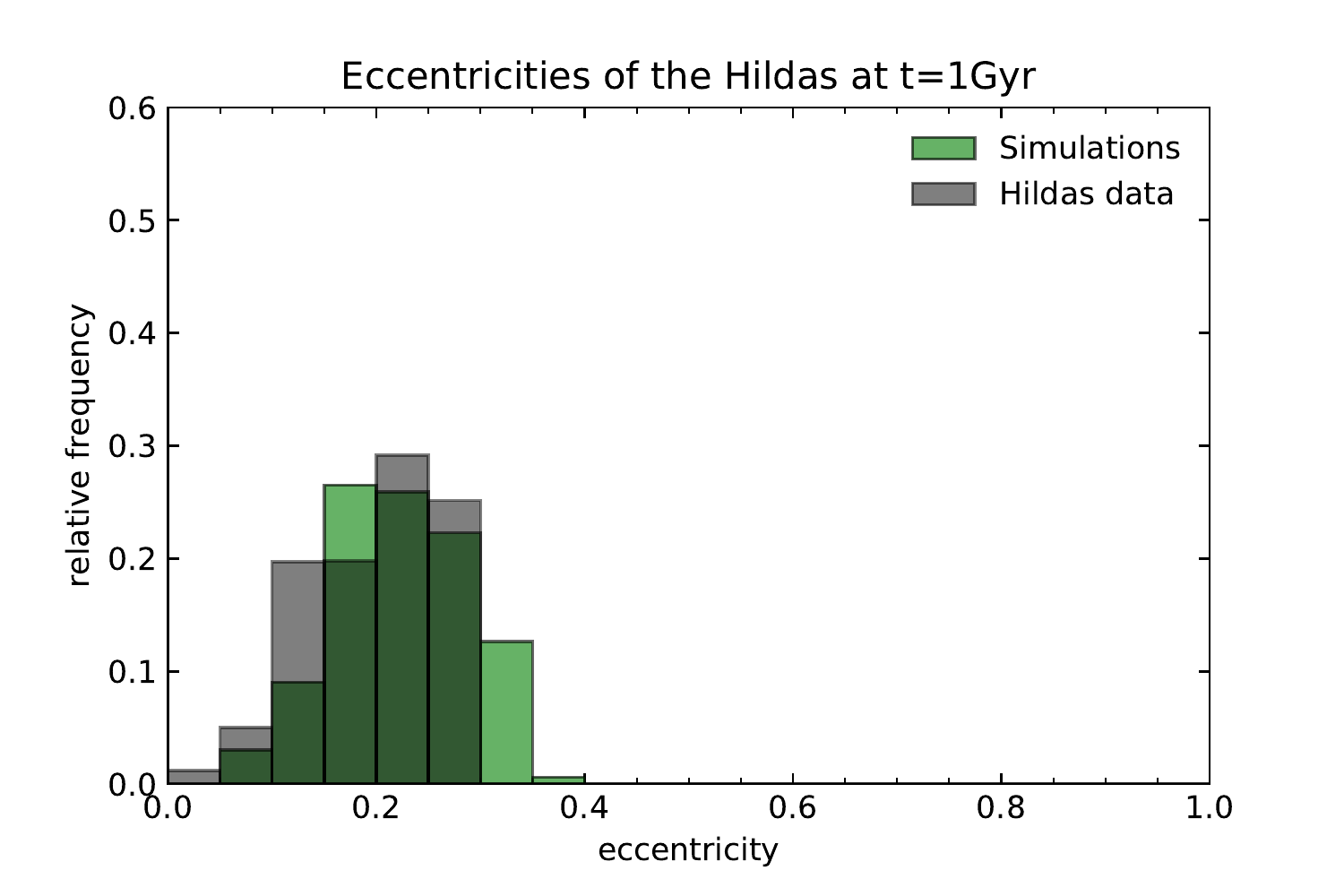}
\includegraphics[width=\hsize]{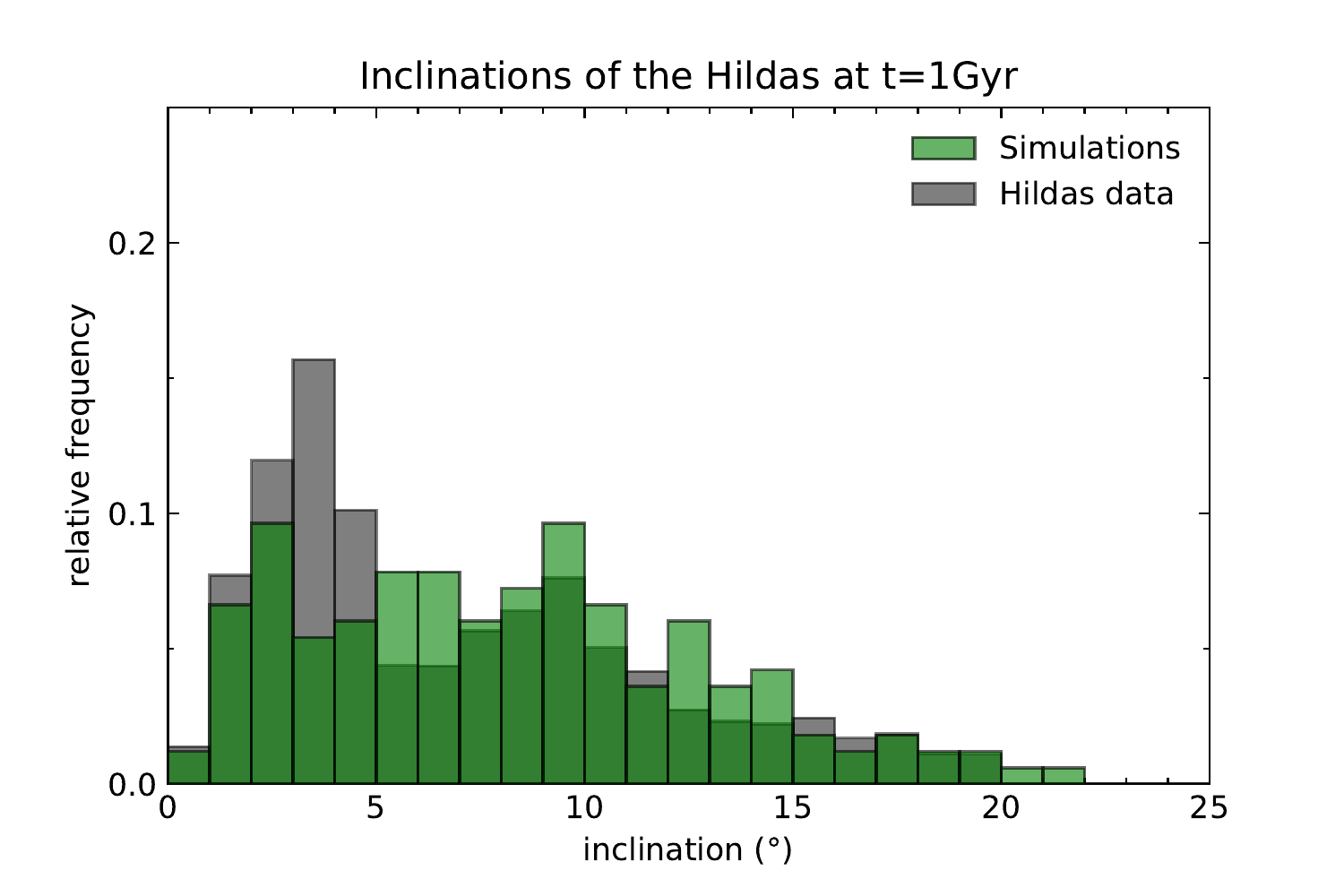}
\caption[]{Osculating eccentricities (top figure) and inclinations (bottom figure) of the Hilda asteroids at $t=1$ Gyr in a model where Jupiter and Saturn, after their growth and early migration where they ended up in the 2:1 resonance, subsequently migrate to their final orbits on a characteristic time scale of $\tau=0.5$ Myr.}
\label{fig:nicehildas}
\end{center}
\end{figure}
The resulting eccentricity and inclination distributions of the Hildas after  $t=1$ Gyr for $\tau=0.5$ Myr is shown in Figure \ref{fig:nicehildas} and they are very similar to the current ones.

\begin{figure}
\begin{center}
\includegraphics[width=\hsize]{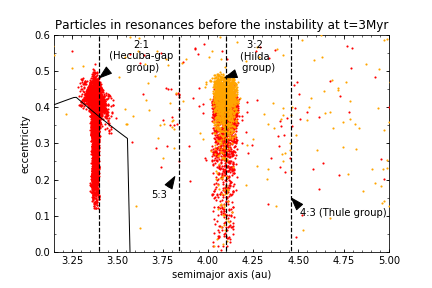}
\includegraphics[width=\hsize]{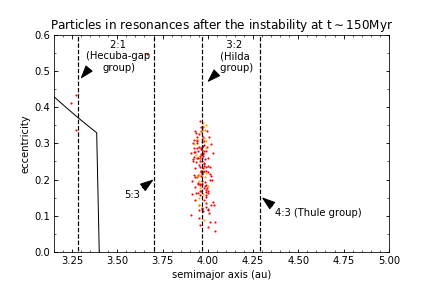}
\caption[]{In the top plot is shown the resonant groups after the early inward migration of the giant planets at $t=3$ Myr. Vertical dashed lines indicate roughly the positions of the resonances and the solid black line delimits the asteroid belt region. In the bottom plot, we show the fate of the resonant populations after the instability of the giant planets with a characteristic time scale of $\tau=0.5$ Myr, at $t\sim150$ Myr. The resonance positions and the asteroid belt region are shifted in the bottom plot because Jupiter migrated from 5.4 to 5.2 au.}
\label{fig:res23}
\end{center}
\end{figure}

We also analysed the survival of the resonant groups we found in Figure \ref{fig:res}. Since in these new simulations, after the early migration, Jupiter end up at 5.4 au we highlighted the new locations of the resonant groups in Figure \ref{fig:res23}, top plot. The asteroid belt region (black line) is properly set based on Jupiter semimajor axis. The 2:1 and 3:2 clumps are very massive as they were in Figure \ref{fig:res}; 5:3 and 4:3 resonant group are not distinguishable in the figure, because at $t=3$ Myr Jupiter has just stopped migrating and the region is not cleared yet from unstable asteroids. The location of the resonances are indicated with dashed vertical lines. Once Jupiter and Saturn migrated following the instability with a characteristic time scale of $\tau=0.5$ Myr, the resonant groups are shown in Figure \ref{fig:res23}, bottom plot. The snapshot has been taken at $t\sim150$ Myr. The figure clearly show that resonant populations are drastically affected by the sweeping of the resonances. 5:3 and 4:3 are completely empty and the 2:1 is almost empty. The Hildas have suffered an important depletion, but the population survives the instability.
In the smooth migration model 6\% of the Hilda survives and in the single large jump of Jupiter's semimajor axis ($\Delta a=0.16$), only 1\% of the Hildas survives.
This finding can be a way to solve the conflict between the fact that nowadays the collision probability for the Jupiter Trojans is higher than for the Hildas, but Trojans
are considered to be less collisionally evolved than Hilda asteroids \citep{wong17,terai18}. From our simulations, after the early migration, we found that Hilda asteroids eccentricities were higher than nowadays and the number of asteroids in the population was larger (Figures \ref{fig:mh0} and \ref{fig:res23}). This imply probably an higher collision probabilities for the Hildas in this phase. After the instability of the giant planets Hildas are severely depleted and eccentricities are lower. In this picture, it is reasonable that nowadays Hilda asteroids are more collisionally evolved than Trojans even if they have lower collision probabilities.


\section{Discussions and conclusions}\label{sec:conclusions}

In this paper, we simulated the mass growth and migration of the giant planets of our solar system according to the core accretion model boosted by pebble accretion in order to allow the cores of the giant planets to grow before the disc disperses. Given the rapid Type I and Type II migration of the forming giant planets through the disc, the planets' seeds were implanted in the outer solar system, with Jupiter's seed starting at about 18 au in our nominal model. We were mainly interested on the effects such large-scale migration could have on the small body populations of our solar system in the first-order resonances with Jupiter, such as the Hilda asteroids and the Trojan asteroids. Our main findings are:
\begin{enumerate}[(a)]

\item After the migration and growth of Jupiter, its Trojan asteroids are characterised by an asymmetry ratio between the leading L$_4$ and the trailing L$_5$ swarms, with the the L$_4$ group always more populated than the L$_5$ group. The asymmetry found is comparable with the one observed in the Jupiter Trojan population. The asymmetry is due to an excess of particles orbiting in the L$_4$ side of the horseshoe orbit while the mass growth of Jupiter shrinks those orbits into stable tadpole orbits. The reason of the excess of particles in the L$_4$ side of the horseshoe orbit is because these orbit are deformed by the relative drift between the migrating planet and the particles in the coorbital resonance \citep{sicardy03,ogilvie06}. In case of inward migration the width of the path of the particles in the L$_4$ side of the horseshoe is wider and hence the particles spend more time there, producing the excess. The in situ growth of the giant planets leads to a symmetric ratio of the number of Trojans between the two swarms. In this case, indeed the horseshoe orbits are symmetric because they are not deformed by any relative drift. Moreover, we found that in order to get an asymmetry ratio comparable to the observed and estimated one, Jupiter needs to migrate inwards at least 3.5 au, that is it must migrate fast in order to get enough relative drift when the core is growing and keep the excess in L$_4$ until the mass growth shrinks the horseshoe orbits.

\item Outward migration while Jupiter is growing leads to an opposite asymmetry ratio between the leading L$_4$ and the trailing L$_5$ swarms compared to the observed one.

\item For the Neptune Trojans we predict no asymmetry. Indeed the mass growth of Neptune in our simulation is not enough to shrink the horseshoe orbits into stable tadpole orbits while there is the excess ahead of the planet. Therefore when the relative drift stops, the horseshoe orbits return to a symmetric geometry and the excess in L$_4$ disappears.

\item Jupiter Trojans are captured from Jupiter's core feeding zone at about 18 au. Neptune Trojans are bodies from Neptune's core feeding zone at about 24 au, but also scattered bodies from the formation locations of the other planets (17-25 au). This is consistent with new results about the colour of the Jupiter Trojans being indistinguishable from the Neptune Trojans, but statistically different from the hot KBOs \citep{jewitt18}.
The Hilda asteroids, instead, are trapped from a region between 5-8 au. Even in this case, this result could be consistent with the similarities of these asteroids with the Jupiter Trojans and the slight difference in their less-red fraction of the populations. The fact that the optical colour distribution of the Jupiter Trojans is indistinguishable from the Neptune Trojans one is a suggestion that Jupiter formed more likely in the outer solar system, closer to the formation region of Neptune than closer to the Hilda asteroids formation region.
In the in situ model, Jupiter Trojans and Hildas are from the same narrow region between 4 and 5 au and this could be in conflict with the discrepancies in their spectra \citep{wong17}.

\item In our scenario explored here, the Jupiter Trojans are a primordial population in which Jupiter's core formed. Therefore, they hold precious information about the building blocks of our giant planets' cores.

\item Jupiter Trojans and their asymmetry survive a possible instability of the giant planets after the disc dispersal, both in case of smooth planetesimal-driven migration and Jumping Jupiter-like migration.

\item The growth and the migration of the giant planets injected bodies formed in the region 4-16 au into the asteroid belt and into the terrestrial planet region and this may have important implications for water delivery to the inner solar system.

\item The eccentricity distribution of Jupiter Trojans is found to be very similar to the observed one, but the inclinations show an almost completely flat distribution, even after we triggered an instability of the giant planets. Their initial values increase of a couple of orders of magnitude, but the distribution remains very flat in contrast with the observed one, where the inclinations are up to 40$^\circ$. The excitation of the Trojan inclinations and the depletion of the swarms will be the topic of our next upcoming publication where we will study the effect of embryos embedded in the swarms and follow the dynamical evolution of the Trojans over billions of years. Since we find in our simulations a very massive Trojan population (3-4 order of magnitude higher than the current one) the primordial Trojans could have contained also massive planetesimals or even planetary embryos ($10^{-4}-10^{-3}$ M$_\oplus$). Such embryos could have been responsible for depleting the swarms and raising the inclinations of the Trojans.

\item The eccentricity and the inclination distributions of the Hilda asteroids completely agree with the current ones after we triggered the dynamical instability. Other populations in resonance with Jupiter, such as 2:1, 5:3 and 4:3 are populated before the primordial large scale migration, but they are emptied during the late instability. Moreover, the fact that before the instability the Hilda group was more populated and more eccentric could 
be an explanation why nowadays Hilda asteroids are more collisionally evolved than Trojans even if they have lower collision probabilities.

\end{enumerate}

The large scale migration predicted in the core accretion model via pebble accretion offers a new possibility to explain the early history of our solar system. Moreover, with NASA's upcoming Lucy mission \citep{levison16} we will have the chance to open a window on the primordial population from which Jupiter's core formed, getting important information about the building material of Jupiter's core that is preserved in the Trojan population.


\begin{acknowledgements}
In loving memory of Angioletta Coradini.
The authors want to thank Pablo Benitez-Llambay for his precious suggestions during the Lund-CPH disc meetings.
Simona Pirani, Anders Johansen and Alexander J. Mustill are supported by the project grant `IMPACT' from the Knut and Alice Wallenberg Foundation (grant 2014.0017). Anders Johansen was further supported by the Knut and Alice Wallenberg Foundation grants 2012.0150 and 2014.0048, the Swedish Research Council (grant 2014-5775) and the European Research Council (ERC Consolidator Grant 724687- PLANETESYS). Bertram Bitsch also thanks the European Research Council (ERC Starting Grant 757448-PAMDORA) for their financial support. The computations are performed on resources provided by the Swedish Infrastructure for Computing (SNIC) at the LUNARC-Centre in Lund.      
\end{acknowledgements}

%
   \bibliographystyle{aa} 
   \bibliography{pirani19} 

\begin{thebibliography}{103}
\expandafter\ifx\csname natexlab\endcsname\relax\def\natexlab#1{#1}\fi

\bibitem[{{Adachi} {et~al.}(1976){Adachi}, {Hayashi}, \& {Nakazawa}}]{adachi76}
{Adachi}, I., {Hayashi}, C., \& {Nakazawa}, K. 1976, Progress of Theoretical
  Physics, 56, 1756

\bibitem[{{Barucci} {et~al.}(2002){Barucci}, {Cruikshank}, {Mottola}, \&
  {Lazzarin}}]{barucci02}
{Barucci}, M.~A., {Cruikshank}, D.~P., {Mottola}, S., \& {Lazzarin}, M. 2002,
  {Physical Properties of Trojan and Centaur Asteroids}, ed. W.~F. {Bottke},
  Jr., A.~{Cellino}, P.~{Paolicchi}, \& R.~P. {Binzel}, 273--287

\bibitem[{{Birnstiel} {et~al.}(2012){Birnstiel}, {Klahr}, \&
  {Ercolano}}]{birnstiel12}
{Birnstiel}, T., {Klahr}, H., \& {Ercolano}, B. 2012, \aap, 539, A148

\bibitem[{{Bitsch} {et~al.}(2015){Bitsch}, {Lambrechts}, \&
  {Johansen}}]{bitsch15b}
{Bitsch}, B., {Lambrechts}, M., \& {Johansen}, A. 2015, \aap, 582, A112

\bibitem[{{Bitsch} {et~al.}(2018){Bitsch}, {Lambrechts}, \&
  {Johansen}}]{bitsch18}
{Bitsch}, B., {Lambrechts}, M., \& {Johansen}, A. 2018, \aap, 609, C2

\bibitem[{{Blum} \& {M{\"u}nch}(1993)}]{blum93}
{Blum}, J. \& {M{\"u}nch}, M. 1993, \icarus, 106, 151

\bibitem[{{Bottke} {et~al.}(2005){Bottke}, {Durda}, {Nesvorn{\'y}}, {Jedicke},
  {Morbidelli}, {Vokrouhlick{\'y}}, \& {Levison}}]{bottke05b}
{Bottke}, W.~F., {Durda}, D.~D., {Nesvorn{\'y}}, D., {et~al.} 2005, \icarus,
  179, 63

\bibitem[{{Brauer} {et~al.}(2008){Brauer}, {Dullemond}, \&
  {Henning}}]{brauer08}
{Brauer}, F., {Dullemond}, C.~P., \& {Henning}, T. 2008, \aap, 480, 859

\bibitem[{{Chambers}(1999)}]{chambers99}
{Chambers}, J.~E. 1999, Monthly Notices of the Royal Astronomical Society, 304,
  793

\bibitem[{{Chambers} \& {Wetherill}(2001)}]{chambers01}
{Chambers}, J.~E. \& {Wetherill}, G.~W. 2001, Meteoritics and Planetary
  Science, 36, 381

\bibitem[{{Cresswell} \& {Nelson}(2008)}]{cresswell08}
{Cresswell}, P. \& {Nelson}, R.~P. 2008, \aap, 482, 677

\bibitem[{{DeMeo} \& {Carry}(2014)}]{demeo14}
{DeMeo}, F.~E. \& {Carry}, B. 2014, \nat, 505, 629

\bibitem[{{Dermott}(1984)}]{dermott84}
{Dermott}, S.~F. 1984, in IAU Colloq. 75: Planetary Rings, ed. R.~{Greenberg}
  \& A.~{Brahic}, 589--637

\bibitem[{{Duncan} {et~al.}(1998){Duncan}, {Levison}, \& {Lee}}]{duncan98}
{Duncan}, M.~J., {Levison}, H.~F., \& {Lee}, M.~H. 1998, The Astronomical
  Journal, 116, 2067

\bibitem[{{Fernandez} \& {Ip}(1984)}]{fernandez84}
{Fernandez}, J.~A. \& {Ip}, W.-H. 1984, \icarus, 58, 109

\bibitem[{{Fleming} \& {Hamilton}(2000)}]{fleming00}
{Fleming}, H.~J. \& {Hamilton}, D.~P. 2000, \icarus, 148, 479

\bibitem[{{Franklin} {et~al.}(1993){Franklin}, {Lecar}, \&
  {Murison}}]{franklin93}
{Franklin}, F., {Lecar}, M., \& {Murison}, M. 1993, \aj, 105, 2336

\bibitem[{{Franklin} {et~al.}(2004){Franklin}, {Lewis}, {Soper}, \&
  {Holman}}]{franklin04}
{Franklin}, F.~A., {Lewis}, N.~K., {Soper}, P.~R., \& {Holman}, M.~J. 2004,
  \aj, 128, 1391

\bibitem[{Gomes {et~al.}(2005)Gomes, Levison, Tsiganis, \&
  Morbidelli}]{gomes05}
Gomes, R., Levison, H.~F., Tsiganis, K., \& Morbidelli, A. 2005, Nature, 435,
  466

\bibitem[{{Gomes}(1998)}]{gomes98}
{Gomes}, R.~S. 1998, \aj, 116, 2590

\bibitem[{{Grav} {et~al.}(2011){Grav}, {Mainzer}, {Bauer}, {Masiero}, {Spahr},
  {McMillan}, {Walker}, {Cutri}, {Wright}, {Eisenhardt}, {Blauvelt}, {DeBaun},
  {Elsbury}, {Gautier}, {Gomillion}, {Hand}, \& {Wilkins}}]{grav11}
{Grav}, T., {Mainzer}, A.~K., {Bauer}, J., {et~al.} 2011, \apj, 742, 40

\bibitem[{{Grazier} {et~al.}(2014){Grazier}, {Castillo-Rogez}, \&
  {Sharp}}]{grazier14}
{Grazier}, K.~R., {Castillo-Rogez}, J.~C., \& {Sharp}, P.~W. 2014, \icarus,
  232, 13

\bibitem[{{G{\"u}ttler} {et~al.}(2010){G{\"u}ttler}, {Blum}, {Zsom}, {Ormel},
  \& {Dullemond}}]{guttler10}
{G{\"u}ttler}, C., {Blum}, J., {Zsom}, A., {Ormel}, C.~W., \& {Dullemond},
  C.~P. 2010, \aap, 513, A56

\bibitem[{{Hartmann} {et~al.}(1998){Hartmann}, {Calvet}, {Gullbring}, \&
  {D'Alessio}}]{hartmann98}
{Hartmann}, L., {Calvet}, N., {Gullbring}, E., \& {D'Alessio}, P. 1998, \apj,
  495, 385

\bibitem[{{Hayashi}(1981)}]{hayashi81}
{Hayashi}, C. 1981, Progress of Theoretical Physics Supplement, 70, 35

\bibitem[{{Ida} {et~al.}(2016){Ida}, {Guillot}, \& {Morbidelli}}]{ida16}
{Ida}, S., {Guillot}, T., \& {Morbidelli}, A. 2016, \aap, 591, A72

\bibitem[{{Jewitt}(2018)}]{jewitt18}
{Jewitt}, D. 2018, \aj, 155, 56

\bibitem[{{Johansen} \& {Lacerda}(2010)}]{johansen10}
{Johansen}, A. \& {Lacerda}, P. 2010, \mnras, 404, 475

\bibitem[{{Johansen} \& {Lambrechts}(2017)}]{johansen17}
{Johansen}, A. \& {Lambrechts}, M. 2017, Annual Review of Earth and Planetary
  Sciences, 45, 359

\bibitem[{{Johansen} {et~al.}(2015){Johansen}, {Mac Low}, {Lacerda}, \&
  {Bizzarro}}]{johansen15}
{Johansen}, A., {Mac Low}, M.-M., {Lacerda}, P., \& {Bizzarro}, M. 2015,
  Science Advances, 1, 1500109

\bibitem[{{Johansen} \& {Youdin}(2007)}]{johansen07}
{Johansen}, A. \& {Youdin}, A. 2007, \apj, 662, 627

\bibitem[{{Kary} \& {Lissauer}(1995)}]{kary95}
{Kary}, D.~M. \& {Lissauer}, J.~J. 1995, \icarus, 117, 1

\bibitem[{{Kley} {et~al.}(2004){Kley}, {Peitz}, \& {Bryden}}]{kley04}
{Kley}, W., {Peitz}, J., \& {Bryden}, G. 2004, \aap, 414, 735

\bibitem[{{Lambrechts} \& {Johansen}(2012)}]{lambrechts12}
{Lambrechts}, M. \& {Johansen}, A. 2012, \aap, 544, A32

\bibitem[{{Lambrechts} \& {Johansen}(2014)}]{lambrechts14b}
{Lambrechts}, M. \& {Johansen}, A. 2014, \aap, 572, A107

\bibitem[{{Lambrechts} {et~al.}(2014){Lambrechts}, {Johansen}, \&
  {Morbidelli}}]{lambrechts14}
{Lambrechts}, M., {Johansen}, A., \& {Morbidelli}, A. 2014, \aap, 572, A35

\bibitem[{{Levison} {et~al.}(2015){Levison}, {Kretke}, \& {Duncan}}]{levison15}
{Levison}, H.~F., {Kretke}, K.~A., \& {Duncan}, M.~J. 2015, \nat, 524, 322

\bibitem[{{Levison} \& {Lucy Science Team}(2016)}]{levison16}
{Levison}, H.~F. \& {Lucy Science Team}. 2016, in Lunar and Planetary Science
  Conference, Vol.~47, Lunar and Planetary Science Conference, 2061

\bibitem[{{Levison} {et~al.}(1997){Levison}, {Shoemaker}, \&
  {Shoemaker}}]{levison97}
{Levison}, H.~F., {Shoemaker}, E.~M., \& {Shoemaker}, C.~S. 1997, \nat, 385, 42

\bibitem[{{Levison} {et~al.}(2010){Levison}, {Thommes}, \&
  {Duncan}}]{levison10}
{Levison}, H.~F., {Thommes}, E., \& {Duncan}, M.~J. 2010, \aj, 139, 1297

\bibitem[{{Levison} {et~al.}(2011){Levison}, {Walsh}, {Barr}, \&
  {Dones}}]{levison11}
{Levison}, H.~F., {Walsh}, K.~J., {Barr}, A.~C., \& {Dones}, L. 2011, Icarus,
  214, 773

\bibitem[{{Lin} \& {Papaloizou}(1986)}]{lin86}
{Lin}, D.~N.~C. \& {Papaloizou}, J. 1986, \apj, 309, 846

\bibitem[{{Lykawka} \& {Horner}(2010)}]{lykawka10}
{Lykawka}, P.~S. \& {Horner}, J. 2010, \mnras, 405, 1375

\bibitem[{{Malhotra}(1993)}]{malhotra93}
{Malhotra}, R. 1993, \nat, 365, 819

\bibitem[{{Malhotra}(1995)}]{malhotra95}
{Malhotra}, R. 1995, \aj, 110, 420

\bibitem[{{Marzari} \& {Scholl}(1998{\natexlab{a}})}]{marzari98a}
{Marzari}, F. \& {Scholl}, H. 1998{\natexlab{a}}, \icarus, 131, 41

\bibitem[{{Marzari} \& {Scholl}(1998{\natexlab{b}})}]{marzari98b}
{Marzari}, F. \& {Scholl}, H. 1998{\natexlab{b}}, \aap, 339, 278

\bibitem[{{Marzari} {et~al.}(2002){Marzari}, {Scholl}, {Murray}, \&
  {Lagerkvist}}]{marzari02}
{Marzari}, F., {Scholl}, H., {Murray}, C., \& {Lagerkvist}, C. 2002, {Origin
  and Evolution of Trojan Asteroids}, ed. W.~F. {Bottke}, Jr., A.~{Cellino},
  P.~{Paolicchi}, \& R.~P. {Binzel}, 725--738

\bibitem[{{Masset} \& {Snellgrove}(2001)}]{masset01}
{Masset}, F. \& {Snellgrove}, M. 2001, \mnras, 320, L55

\bibitem[{{Minton} \& {Malhotra}(2009)}]{minton09}
{Minton}, D.~A. \& {Malhotra}, R. 2009, Nature, 457, 1109

\bibitem[{{Minton} \& {Malhotra}(2010)}]{minton10}
{Minton}, D.~A. \& {Malhotra}, R. 2010, Icarus, 207, 744

\bibitem[{{Morbidelli} {et~al.}(2010){Morbidelli}, {Brasser}, {Gomes},
  {Levison}, \& {Tsiganis}}]{morbidelli10}
{Morbidelli}, A., {Brasser}, R., {Gomes}, R., {Levison}, H.~F., \& {Tsiganis},
  K. 2010, The Astronomical Journal, 140, 1391

\bibitem[{Morbidelli {et~al.}(2005)Morbidelli, Levison, Tsiganis, \&
  Gomes}]{morbidelli05}
Morbidelli, A., Levison, H.~F., Tsiganis, K., \& Gomes, R. 2005, Nature, 435,
  462

\bibitem[{{Morbidelli} {et~al.}(2007){Morbidelli}, {Tsiganis}, {Crida},
  {Levison}, \& {Gomes}}]{morbidelli07}
{Morbidelli}, A., {Tsiganis}, K., {Crida}, A., {Levison}, H.~F., \& {Gomes}, R.
  2007, Astronomical Journal, 134, 1790

\bibitem[{{Murray}(1994)}]{murray94}
{Murray}, C.~D. 1994, \icarus, 112, 465

\bibitem[{{Murray} \& {Dermott}(1999)}]{murray99}
{Murray}, C.~D. \& {Dermott}, S.~F. 1999, {Solar system dynamics}

\bibitem[{{Nakamura} \& {Yoshida}(2008)}]{nakamura2008b}
{Nakamura}, T. \& {Yoshida}, F. 2008, \pasj, 60, 293

\bibitem[{{Nesvorn{\'y}}(2011)}]{nesvorny11}
{Nesvorn{\'y}}, D. 2011, \apjl, 742, L22

\bibitem[{{Nesvorn{\'y}} {et~al.}(2013){Nesvorn{\'y}}, {Vokrouhlick{\'y}}, \&
  {Morbidelli}}]{nesvorny13}
{Nesvorn{\'y}}, D., {Vokrouhlick{\'y}}, D., \& {Morbidelli}, A. 2013, \apj,
  768, 45

\bibitem[{{O'Brien} {et~al.}(2007){O'Brien}, {Morbidelli}, \&
  {Bottke}}]{obrien07}
{O'Brien}, D.~P., {Morbidelli}, A., \& {Bottke}, W.~F. 2007, \icarus, 191, 434

\bibitem[{{Ogilvie} \& {Lubow}(2006)}]{ogilvie06}
{Ogilvie}, G.~I. \& {Lubow}, S.~H. 2006, \mnras, 370, 784

\bibitem[{{Okuzumi} {et~al.}(2012){Okuzumi}, {Tanaka}, {Kobayashi}, \&
  {Wada}}]{okuzumi12}
{Okuzumi}, S., {Tanaka}, H., {Kobayashi}, H., \& {Wada}, K. 2012, \apj, 752,
  106

\bibitem[{{Ormel} \& {Klahr}(2010)}]{ormel10}
{Ormel}, C.~W. \& {Klahr}, H.~H. 2010, \aap, 520, A43

\bibitem[{{Peale}(1993)}]{peale93}
{Peale}, S.~J. 1993, \icarus, 106, 308

\bibitem[{{Petit} {et~al.}(2001){Petit}, {Morbidelli}, \& {Chambers}}]{petit01}
{Petit}, J.-M., {Morbidelli}, A., \& {Chambers}, J. 2001, \icarus, 153, 338

\bibitem[{{Petit} {et~al.}(1998){Petit}, {Morbidelli}, \&
  {Valsecchi}}]{petit98}
{Petit}, J.-M., {Morbidelli}, A., \& {Valsecchi}, G. 1998, in \baas, Vol.~30,
  Bulletin of the American Astronomical Society, 1453

\bibitem[{{Pierens} \& {Nelson}(2008)}]{pierens08}
{Pierens}, A. \& {Nelson}, R.~P. 2008, \aap, 482, 333

\bibitem[{{Pirani} \& {Turrini}(2016)}]{pirani16}
{Pirani}, S. \& {Turrini}, D. 2016, \icarus, 271, 170

\bibitem[{{Pollack} {et~al.}(1996){Pollack}, {Hubickyj}, {Bodenheimer},
  {Lissauer}, {Podolak}, \& {Greenzweig}}]{pollack96}
{Pollack}, J.~B., {Hubickyj}, O., {Bodenheimer}, P., {et~al.} 1996, \icarus,
  124, 62

\bibitem[{{Rafikov}(2004)}]{rafikov04}
{Rafikov}, R.~R. 2004, \aj, 128, 1348

\bibitem[{{Raymond} \& {Izidoro}(2017{\natexlab{a}})}]{raymond17}
{Raymond}, S.~N. \& {Izidoro}, A. 2017{\natexlab{a}}, \icarus, 297, 134

\bibitem[{{Raymond} \& {Izidoro}(2017{\natexlab{b}})}]{raymond17b}
{Raymond}, S.~N. \& {Izidoro}, A. 2017{\natexlab{b}}, Science Advances, 3,
  e1701138

\bibitem[{{Roig} \& {Nesvorn{\'y}}(2015)}]{roig15}
{Roig}, F. \& {Nesvorn{\'y}}, D. 2015, \aj, 150, 186

\bibitem[{{Ros} \& {Johansen}(2013)}]{ros13}
{Ros}, K. \& {Johansen}, A. 2013, \aap, 552, A137

\bibitem[{{Schoonenberg} \& {Ormel}(2017)}]{schoonenberg17}
{Schoonenberg}, D. \& {Ormel}, C.~W. 2017, \aap, 602, A21

\bibitem[{{Shoemaker} {et~al.}(1989){Shoemaker}, {Shoemaker}, \&
  {Wolfe}}]{shoemaker89}
{Shoemaker}, E.~M., {Shoemaker}, C.~S., \& {Wolfe}, R.~F. 1989, in Asteroids
  II, ed. R.~P. {Binzel}, T.~{Gehrels}, \& M.~S. {Matthews}, 487--523

\bibitem[{{Sicardy} \& {Dubois}(2003)}]{sicardy03}
{Sicardy}, B. \& {Dubois}, V. 2003, Celestial Mechanics and Dynamical
  Astronomy, 86, 321

\bibitem[{{Simon} {et~al.}(2016){Simon}, {Armitage}, {Li}, \&
  {Youdin}}]{simon16}
{Simon}, J.~B., {Armitage}, P.~J., {Li}, R., \& {Youdin}, A.~N. 2016, \apj,
  822, 55

\bibitem[{{Slyusarev} \& {Belskaya}(2014)}]{slyusarev14}
{Slyusarev}, I.~G. \& {Belskaya}, I.~N. 2014, Solar System Research, 48, 139

\bibitem[{{Szab{\'o}} {et~al.}(2007){Szab{\'o}}, {Ivezi{\'c}}, {Juri{\'c}}, \&
  {Lupton}}]{szabo07}
{Szab{\'o}}, G.~M., {Ivezi{\'c}}, {\v Z}., {Juri{\'c}}, M., \& {Lupton}, R.
  2007, \mnras, 377, 1393

\bibitem[{{Tanaka} \& {Ward}(2004)}]{tanaka04}
{Tanaka}, H. \& {Ward}, W.~R. 2004, \apj, 602, 388

\bibitem[{{Terai} \& {Yoshida}(2018)}]{terai18}
{Terai}, T. \& {Yoshida}, F. 2018, ArXiv e-prints [\eprint[arXiv]{1805.09445}]

\bibitem[{{Testi} {et~al.}(2003){Testi}, {Natta}, {Shepherd}, \&
  {Wilner}}]{testi03}
{Testi}, L., {Natta}, A., {Shepherd}, D.~S., \& {Wilner}, D.~J. 2003, \aap,
  403, 323

\bibitem[{Tsiganis {et~al.}(2005)Tsiganis, Gomes, Morbidelli, \&
  Levison}]{tsiganis05}
Tsiganis, K., Gomes, R., Morbidelli, A., \& Levison, H.~F. 2005, Nature, 435,
  459

\bibitem[{Turrini {et~al.}(2011)Turrini, Magni, \& Coradini}]{turrini11}
Turrini, D., Magni, G., \& Coradini, A. 2011, Monthly Notices of the Royal
  Astronomical Society, 413, 2439

\bibitem[{{Turrini} \& {Svetsov}(2014)}]{turrini14b}
{Turrini}, D. \& {Svetsov}, V. 2014, Life, 4, 4

\bibitem[{{Venturini} \& {Helled}(2017)}]{venturini17}
{Venturini}, J. \& {Helled}, R. 2017, \apj, 848, 95

\bibitem[{{Vinogradova} \& {Chernetenko}(2015)}]{vinogradova15}
{Vinogradova}, T.~A. \& {Chernetenko}, Y.~A. 2015, Solar System Research, 49,
  391

\bibitem[{{Vokrouhlick{\'y}} {et~al.}(2016){Vokrouhlick{\'y}}, {Bottke}, \&
  {Nesvorn{\'y}}}]{vokrouhlicky16}
{Vokrouhlick{\'y}}, D., {Bottke}, W.~F., \& {Nesvorn{\'y}}, D. 2016, \aj, 152,
  39

\bibitem[{{Wada} {et~al.}(2009){Wada}, {Tanaka}, {Suyama}, {Kimura}, \&
  {Yamamoto}}]{wada09}
{Wada}, K., {Tanaka}, H., {Suyama}, T., {Kimura}, H., \& {Yamamoto}, T. 2009,
  \apj, 702, 1490

\bibitem[{Walsh {et~al.}(2011)Walsh, Morbidelli, Raymond, O'Brien, \&
  Mandell}]{walsh11}
Walsh, K.~J., Morbidelli, A., Raymond, S.~N., O'Brien, D.~P., \& Mandell, A.~M.
  2011, Nature, 475, 206

\bibitem[{{Ward}(1997)}]{ward97}
{Ward}, W.~R. 1997, \icarus, 126, 261

\bibitem[{{Weidenschilling}(1977)}]{weidenschilling77}
{Weidenschilling}, S.~J. 1977, \mnras, 180, 57

\bibitem[{Wetherill(1992)}]{wetherill92}
Wetherill, G.~W. 1992, Icarus, 100, 307

\bibitem[{{Whipple}(1972)}]{whipple72}
{Whipple}, F.~L. 1972, in From Plasma to Planet, ed. A.~{Elvius}, 211

\bibitem[{{Wilner} {et~al.}(2005){Wilner}, {D'Alessio}, {Calvet}, {Claussen},
  \& {Hartmann}}]{wilner05}
{Wilner}, D.~J., {D'Alessio}, P., {Calvet}, N., {Claussen}, M.~J., \&
  {Hartmann}, L. 2005, \apjl, 626, L109

\bibitem[{{Windmark} {et~al.}(2012{\natexlab{a}}){Windmark}, {Birnstiel},
  {G{\"u}ttler}, {Blum}, {Dullemond}, \& {Henning}}]{windmark12a}
{Windmark}, F., {Birnstiel}, T., {G{\"u}ttler}, C., {et~al.}
  2012{\natexlab{a}}, \aap, 540, A73

\bibitem[{{Windmark} {et~al.}(2012{\natexlab{b}}){Windmark}, {Birnstiel},
  {Ormel}, \& {Dullemond}}]{windmark12b}
{Windmark}, F., {Birnstiel}, T., {Ormel}, C.~W., \& {Dullemond}, C.~P.
  2012{\natexlab{b}}, \aap, 544, L16

\bibitem[{{Wisdom} \& {Holman}(1991)}]{wisdom91}
{Wisdom}, J. \& {Holman}, M. 1991, The Astronomical Journal, 102, 1528

\bibitem[{{Wong} \& {Brown}(2017)}]{wong17}
{Wong}, I. \& {Brown}, M.~E. 2017, \aj, 153, 69

\bibitem[{{Yoder}(1979)}]{yoder79}
{Yoder}, C.~F. 1979, \icarus, 40, 341

\bibitem[{{Youdin} \& {Goodman}(2005)}]{youdin05}
{Youdin}, A.~N. \& {Goodman}, J. 2005, \apj, 620, 459

\bibitem[{{Zsom} {et~al.}(2010){Zsom}, {Ormel}, {G{\"u}ttler}, {Blum}, \&
  {Dullemond}}]{zsom10}
{Zsom}, A., {Ormel}, C.~W., {G{\"u}ttler}, C., {Blum}, J., \& {Dullemond},
  C.~P. 2010, \aap, 513, A57

\end{thebibliography}
%
\begin{appendix}

\section{Aerodynamic gas drag}\label{sec:agd}

The aerodynamic gas drag is the frictional force between gas and planetesimals. 
In our simulations we wanted to mimic the presence of a gaseous protoplanetary disc by modifying the \textsc{Mercury} \textit{N}-body code to include the effects of the gas drag.

As defined in \citet{adachi76}, the drag force is 
\begin{equation}
F=\rm{A}\rm{\rho_{gas}}\varv_{\rm{rel}}^2
\end{equation}
where $\rho_{\rm{gas}}$ is the volume gas density, $\varv_{\rm{rel}}$ is the relative velocity between a planetesimal and the gas. A is a function of the dimensionless drag coefficient (C$_{\rm{D}}$) and the radius and the mass of the planetesimal ($r_{\rm{p}}$ and $m_{\rm{p}}$):  
\begin{equation}
\rm{A}=\frac{C_{\rm{D}}\pi r_{\rm{p}}^2}{2 m_{\rm{p}}}
\end{equation} 
The characteristic timescale for the drag force is 
\begin{equation}
\tau_{\rm{0}}= \frac{1}{A\rho_{\rm{gas}}\varv_{\rm{kepl}}}=\frac{m_{\rm{p}}}{\pi C_{\rm{D}} r_{\rm{p}}^2 \rho_{\rm{gas}}\varv_{\rm{kepl}}}=\frac{8\rho_{\rm{p}} r_{\rm{p}}}{3 C_{\rm{D}} \rho_{\rm{gas}} \varv_{\rm{kep}}}
\end{equation} where $\rho_{\rm{p}}$ is the density of the planetesimal, $\varv_{\rm{kep}}$ is the keplerian orbital velocity. 

In general, C$_{\rm{D}}$ depends on the ratio between the mean free path of the gas atoms or molecules (L) and the radius of the planetesimal, called Knudsen number (K), on the relative velocity of the particle compared to the sound speed, that is the Mach number, (M) and on the Reynolds number, defined as 
\begin{equation}
\rm{R}_{\rm{e}}=\frac{2\rho_{\rm{gas}} \varv_{\rm{rel}} r_{\rm{p}}}{\nu}
\end{equation} 
where $\nu$ is the viscosity of the gas.

In our simulations, we assign to all the small bodies involved a radius $r_{\rm{p}}=50$ km 
and a density $\rho_{\rm{p}}=1.0$ g/cm$^3$. 
In this limit, where L$< 2 r_{\rm{p}}$ and R$_{\rm{e}}>10^3$, C$_{\rm{D}}$ is not a function of any of K, M and R$_{\rm{e}}$ and has a constant value of 0.44 \citep{whipple72}.

The approximated formulas found by \citet{adachi76} of the time evolution (averaged over a Keplerian period) of the eccentricity $e$, the inclination $i$ and the semimajor axis $a$, are:

\begin{equation}\label{eq:e}
\left.\frac{\dot e}{e}\right|_{\rm{drag}}= -\frac{1}{\tau_{\rm{0}}}\left(\eta^2+\frac{5}{8}e^2+\frac{1}{2}i^2\right)^{1/2}
\end{equation}

\begin{equation}\label{eq:i}
\left.\frac{\dot{i}}{i}\right|_{\rm{drag}}= -\frac{1}{2\tau_{\rm{0}}}\left(\eta^2+\frac{5}{8}e^2+\frac{1}{2}i^2\right)^{1/2}
\end{equation}

\begin{equation}\label{eq:a}
\left.\frac{\dot{a}}{a}\right|_{\rm{drag}}= -\frac{2}{\tau_{\rm{0}}}\left(\eta^2+\frac{5}{8}e^2+\frac{1}{2}i^2\right)^{1/2} \left(\eta+e^2+\frac{1}{8}i^2\right)
\end{equation}

$\eta$ represents the deviation of the gas velocity $\varv_{\rm{gas}}$ from the Keplerian velocity $\varv_{\rm{kep}}$ due to the radial pressure gradient in the gaseous disc and is defined as:
\begin{equation}
\eta = -\frac{\varv_{\rm{rel}}}{\varv_{\rm{kep}}}=\frac{1}{2}\left(\frac{H}{r}\right)^2 \frac{\partial \ln P}{\partial \ln{r}}
\end{equation}

where $H/r=c_{\rm{s}} /\varv_{\rm{kep}}$ is the aspect ratio and $\partial\ln P / \partial\ln{r}$ is the pressure gradient of the disc. 

The aerodynamic gas drag affects our small particles until the gaseous disc dissipates, that is until $t=3$ Myr.


\section{Tidal gas drag}\label{sec:tgd}

The tidal gas drag is the loss of orbital energy due to angular momentum transfer with the protoplanetary disc and affects bodies that are large enough to create density enhancements in the gas of the disc, but are below the mass limit required to initiate large scale gas accretion. In our simulations, we modified the \textsc{Mercury} \textit{N}-body code such as it includes the tidal gas drag effects on the giant planets cores.

The eccentricity and the inclination damping, due to the tidal gas drag, are given by these fits to the simulations in \citet{cresswell08}:
\begin{equation}\begin{split}
\left.\frac{\dot e}{e}\right|_{\rm{tidal}}= & -\frac{0.78}{t_{\rm{wave}}}\Biggr[1-0.14\left(\frac{e}{H/r}\right)^2+0.06\left(\frac{e}{H/r}\right)^3+ \\
& + 0.18\left(\frac{i}{H/r}\right)^2\left(\frac{e}{H/r}\right)\Biggr]^{-1}
\end{split}
\end{equation}

\begin{equation}\begin{split}
\left.\frac{\dot{i}}{i}\right|_{\rm{tidal}}= & -\frac{0.544}{t_{\rm{wave}}}\Biggr[1-0.30\left(\frac{i}{H/r}\right)^2+0.24\left(\frac{i}{H/r}\right)^3+ \\
& + 0.14\left(\frac{e}{H/r}\right)^2\left(\frac{i}{H/r}\right)\Biggr]^{-1}
\end{split}
\end{equation}

where $e$ and $i$ are respectively the eccentricity and the inclination of the planetary cores, $H/r$ is the local disc aspect ratio and $t_{wave}$ is the damping time scale derived by \citet{tanaka04}:

\begin{equation}
t_{\rm{wave}}=\frac{M_*^2}{\Omega_{\rm{c}} m_{\rm{c}} \Sigma_{\rm{c}} a_{\rm{c}}^2} \left( \frac{H}{r}\right)^4
\end{equation}

where M$_*$ is the mass of the central star, $m_{\rm{c}}$ and $a_{\rm{c}}$ are the mass and the semimajor axis of the planetary core, $\Sigma_{\rm{c}}$ is the local disc surface density and $\Omega_{\rm{c}}$ is the orbital angular velocity.

In our simulations, the tidal gas drag affects our protoplanets while they have a mass below $30$ M$_\oplus$ and until the gaseous disc dissipates, that is until $t=3$ Myr.

\end{appendix}

\end{document}